\pdfoutput=1

\documentclass[11pt]{article}

\usepackage{acl}

\usepackage{times}
\usepackage{latexsym}

\usepackage[T1]{fontenc}

\usepackage[utf8]{inputenc}

\usepackage{microtype}

\usepackage{hyperref}       
\usepackage{url}            
\usepackage{booktabs}       
\usepackage{nicefrac}       
\usepackage{microtype}      
\usepackage{xcolor}         
\usepackage{times}  
\usepackage{helvet}  
\usepackage{courier}  
\usepackage{graphicx}  
\usepackage{makecell}
\usepackage{bm}
\usepackage{subcaption}
\usepackage{amsmath,amsfonts,amssymb,amsthm}

\DeclareMathOperator*{\argmin}{arg\,min}

\makeatletter
\newcommand{\specificthanks}[1]{\@fnsymbol{#1}}
\makeatother

\usepackage[export]{adjustbox}
\DeclareMathOperator{\bbE}{\mathbb{E}}
\usepackage{mathtools}
\usepackage{tikz}
\usetikzlibrary{arrows, shapes, decorations.pathreplacing, calc, positioning}
\usepackage{algorithm}
\usepackage{algpseudocode}
\usepackage{fancyvrb}
\usepackage{enumitem,kantlipsum}
\usepackage{tabularx}
\usepackage{multirow}

\title{MICO: Selective Search with Mutual Information Co-training}

\author{Zhanyu Wang\thanks{\quad Work done while doing an internship at Amazon.}
                    \textsuperscript{, \specificthanks{2}}\\
  Purdue University \\
  {\tt \{wang4094\}@purdue.edu} \\
  \AND
  Xiao Zhang\thanks{\quad Equal contribution.} \and Hyokun Yun \and Choon Hui Teo \and Trishul Chilimbi\\
  Amazon \\
  {\tt \{zhxao, yunhyoku, choonhui, trishulc\}@amazon.com} \\}

\begin{document}

\maketitle

\begin{abstract}


In contrast to traditional exhaustive search, selective search first clusters documents into several groups before all the documents are searched exhaustively by a query, to limit the search executed within one group or only a few groups. Selective search is designed to reduce the latency and computation in modern large-scale search systems. In this study, we propose MICO, a \textbf{M}utual \textbf{I}nformation \textbf{CO}-training framework for selective search with minimal supervision using the search logs. After training, MICO does not only cluster the documents, but also routes unseen queries to the relevant clusters for efficient retrieval. In our empirical experiments, MICO significantly improves the performance on multiple metrics of selective search and outperforms a number of existing competitive baselines.

\end{abstract}

\section{Introduction}
\label{sec:intro}



\begin{figure}[h]
\centering
\includegraphics[width=0.50\textwidth, clip]{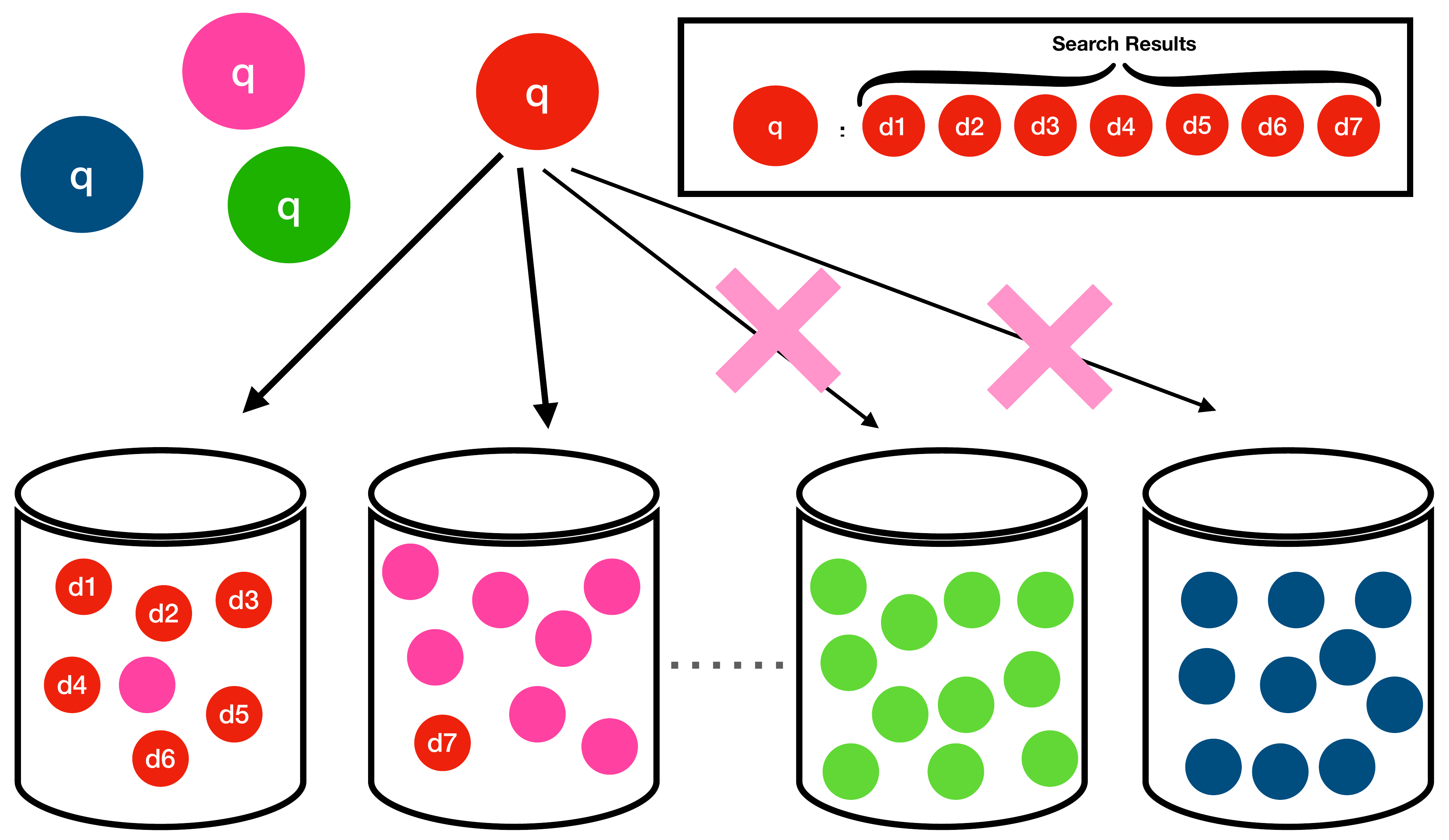}
\caption{Machine learning based selective search first divides documents into different shards based on their semantic similarities and then routes queries to the shards containing documents most relevant to this query.}
\label{fig:selective_search}
\end{figure}

In information retrieval (IR), searching over all the documents is quite costly at a large scale \citep{risvik2013maguro}. Selective search \citep{kulkarni_efficient_2013, kulkarni_selective_2015} divides documents into a number of shards (clusters) to allow an incoming query to search over only a small number of these shards that are most relevant to this query. It tries to preserve the quality of search results similar to that of searching over the whole corpus, with less computation and shorter retrieval latency, achieved by limiting an incoming query to search within only a few chosen document shards. Sharding is the process of breaking a massive collection of documents into smaller chunks called shards. 
Randomly assigning documents to shards, a popular approach, cannot guarantee documents relevant to a specific query restrained within one shard or only a few shards.
With the observation that semantically similar documents tend to appear together in the search results of the same query \citep{van1979information}, researchers in recent years have devised multiple machine learning-based algorithms for selective search. 
These algorithms are tailored to benefit the document retrieval process by dividing a large volume of the corpus in a certain way so that semantically related documents are allocated in the same cluster and then routing new queries to the most relevant clusters subsequently. In this way, the retrieval system can return the search results with both efficiency and accuracy, as shown in Figure \ref{fig:selective_search}.
These approaches significantly differ from the traditional sharding methods that either randomly split the collection into several groups \citep{barroso2003web} or allocate them based on simple handcrafted rules \citep{gravano-1999-gloss}, and have achieved comparable performance to exhaustive search over the entire document corpus.

Nevertheless, having led giant strides in selective search, these approaches fail to address a few critical practical issues when scalable IR systems for a large number of documents in the  industry are needed. A major concern is that these approaches heavily rely on a multi-stage process: firstly, complex clustering procedures are applied to allocate documents into different shards, while later additional elaborate algorithms are used to choose a small number of shards for incoming queries. The multi-stage process needs not only additional infrastructure support to bridge the output of the previous step as the input of the next step but also is prone to error propagation. A natural attempt to mitigate the aforementioned difficulties is to construct a model for both document allocation and subsequent query routing, with an objective function to be optimized in an end-to-end paradigm. This approach ensures applying gradient-based learning to modules or components of the system coherently. 

To this end, we introduce Mutual Information CO-training (MICO), a novel end-to-end approach for selective search. 
We argue that by maximizing the mutual information of the query routing variable and document assignment variable of a relevant query-document pair, we can leverage co-training in an end-to-end training fashion to overcome the aforementioned difficulties. This approach first creates a bipartite graph by exploiting the query-document relations and then treats query and document as two different views of the same example while maximizing the mutual information between the cluster variable of these two views. We show an overview of the MICO framework in Figure \ref{fig:modeling} and summarize our contributions:

\begin{figure*}[ht]
\centering
\begin{subfigure}[b]{0.45\textwidth}
\centering
\includegraphics[width=0.65\textwidth, trim=0 70 340 5, clip]{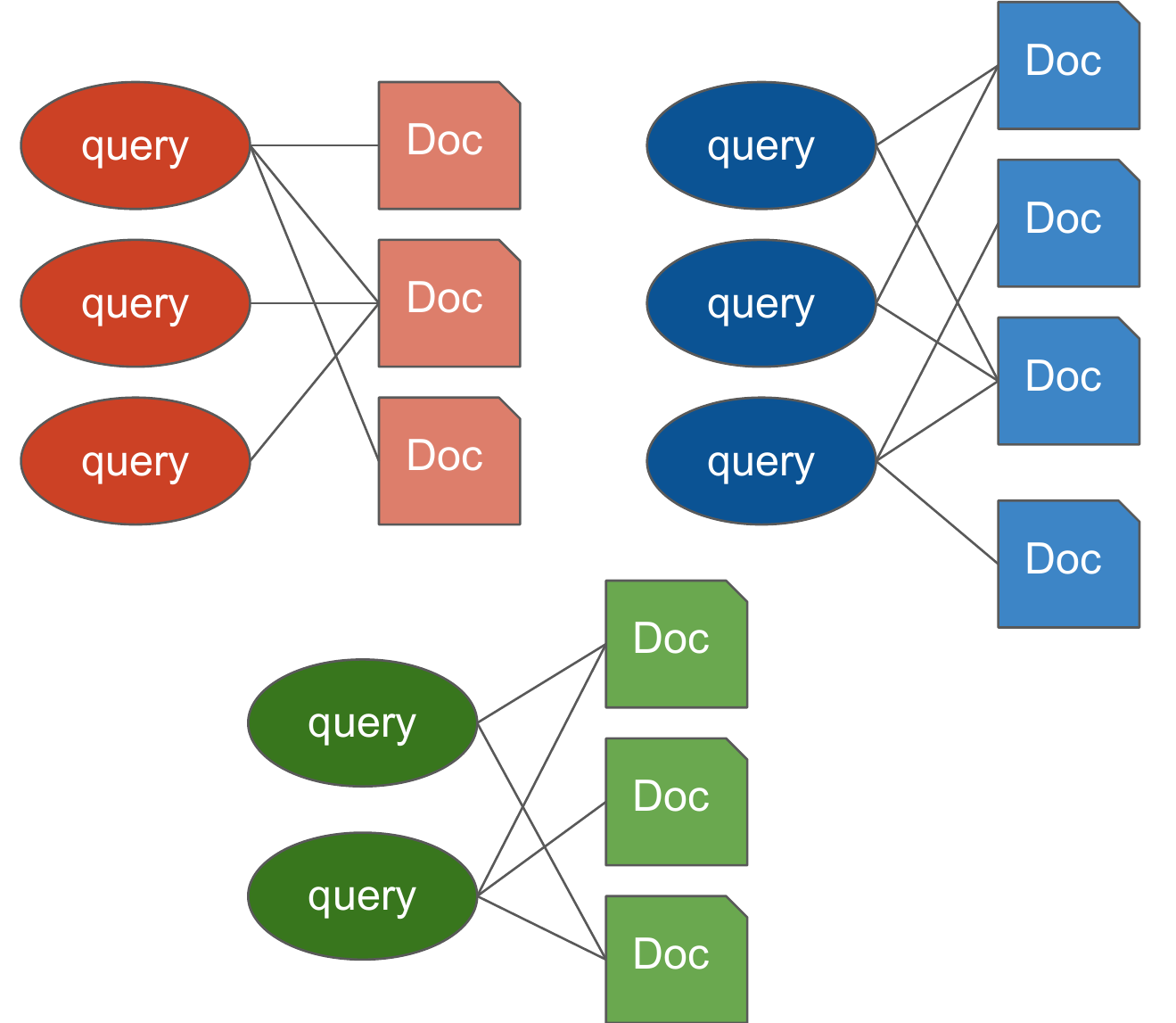}
\caption{Imaginary Document Clusters (with Queries)}
\label{fig:clusters}
\end{subfigure}
\begin{subfigure}[b]{0.53\textwidth}
\centering
\includegraphics[width=0.65\textwidth, trim=0 450 1100 0, clip]{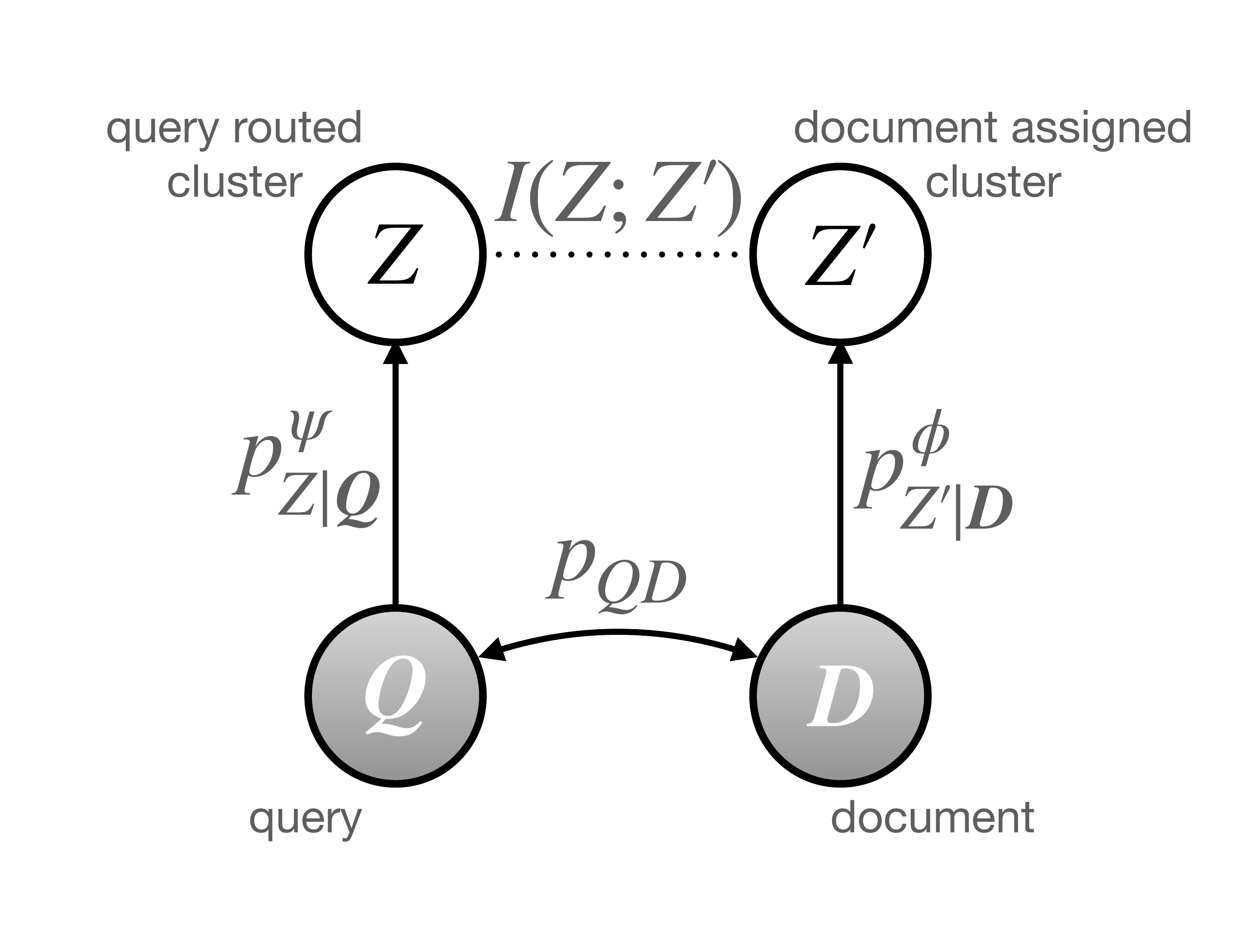}
\caption{Overview of Mutual Information CO-training (MICO)}
\label{fig:mico}
\end{subfigure}
\begin{subfigure}[b]{0.45\textwidth}
\centering
\includegraphics[width=0.7\textwidth, trim=0 70 340 5, clip]{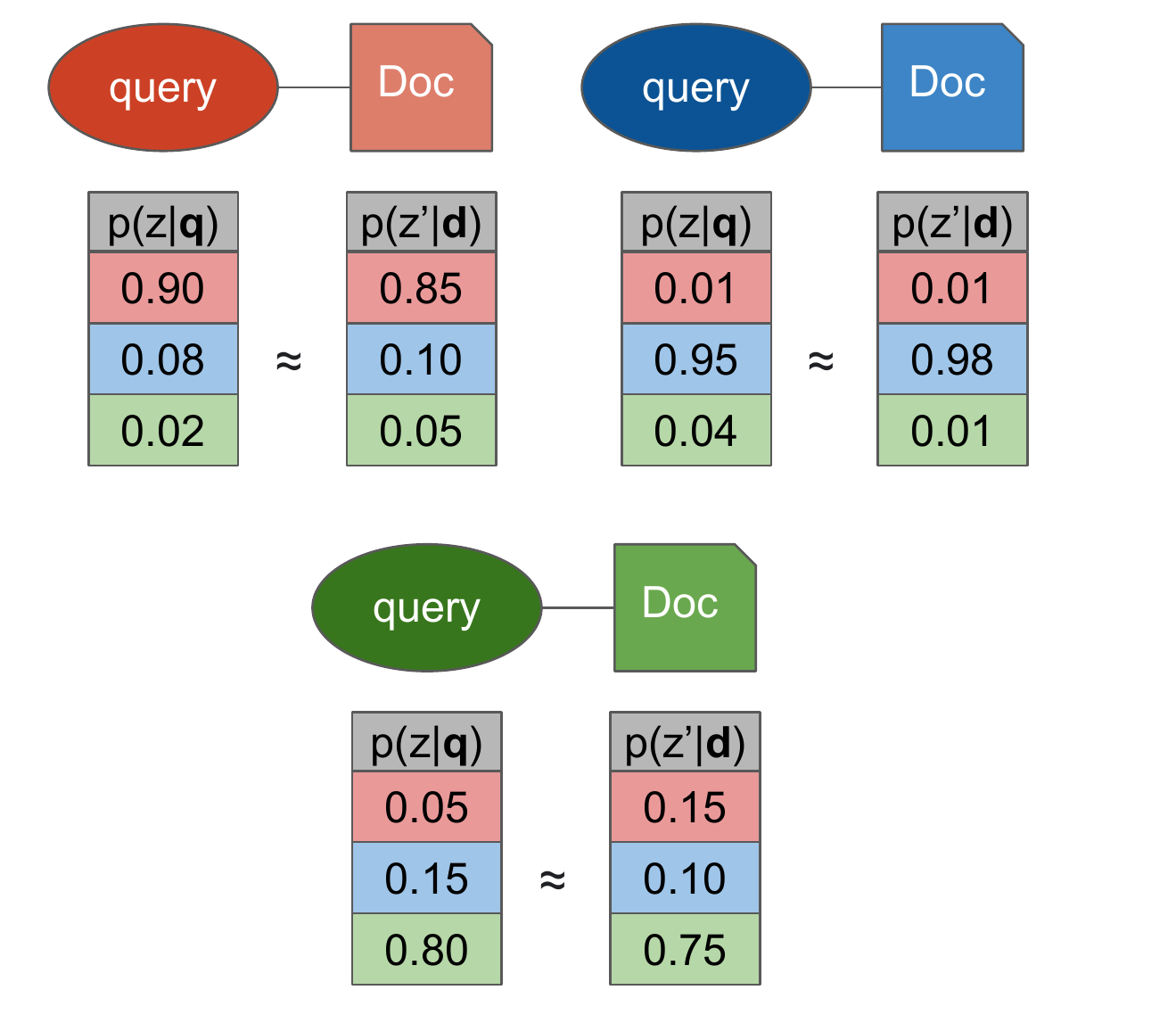}
\caption{Cluster selection coherency through maximizing the mutual information $I(Z;Z')$}
\label{fig:max_mi}
\end{subfigure}
\begin{subfigure}[b]{0.53\textwidth}
\centering
\includegraphics[width=0.4\textwidth, trim=0 70 480 5, clip]{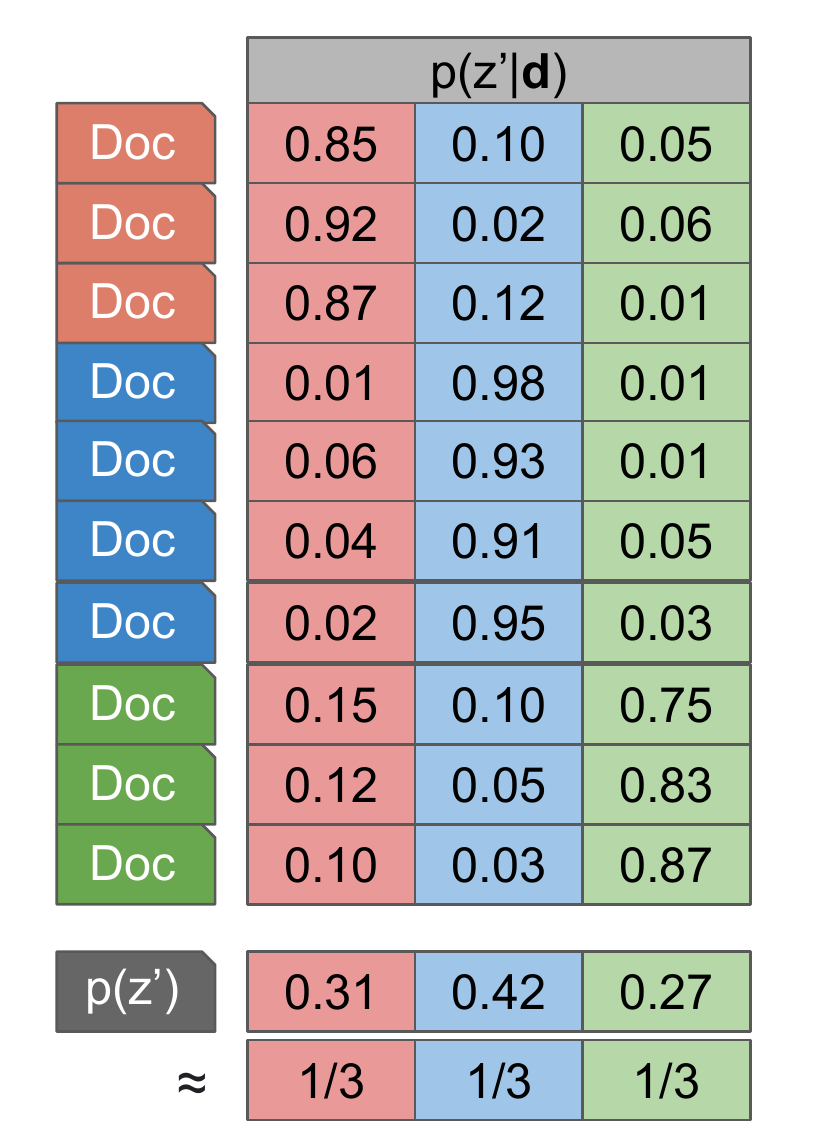}
\caption{Balance of shard size by maximizing the entropy $H(Z')$}
\label{fig:max_entropy}
\end{subfigure}
\caption{Figure \ref{fig:clusters} on the left illustrates the query-document pairs with the clusters they belong to, denoted by red, blue, and green, respectively. Figure \ref{fig:mico} on the right is an overview of our Mutual Information CO-training (MICO) approach: the query and the document are treated as two views of the same example, and two different distributions parameterize the cluster assignment variables of the two views respectively. In Figure \ref{fig:max_mi} and Figure \ref{fig:max_entropy}, we illustrate cluster selection with an example where the predicted cluster index distribution is a vector for each query and the document.}
\label{fig:modeling}
\end{figure*}

\begin{enumerate}[noitemsep, leftmargin=*]
\item We propose MICO, Mutual Information CO-training, a novel approach for selective search. MICO treats the query and the document as two different views of the same training example: the query-document pair in the log, maximizing the mutual information between the cluster assignment variables of each view. The cluster assignment variables are parameterized via two separate distributions conditioned on the query and the document.
\item We design MICO ready for practical use as it is being trained end-to-end for both document sharding and subsequent query routing. While the trained document allocation module assigns documents into different clusters, the trained query routing module routes new queries to the target cluster(s) to retrieve the most relevant documents at a low cost. To the best of our knowledge, MICO is the first attempt to deal with document sharding and query routing in selective search in an end-to-end fashion.
\item We show significantly improved performance on two IR data sets with MICO on multiple important metrics empirically in selective search. MICO beats competitive baselines regarding query coverage while minimizing search costs.
\end{enumerate}

\section{Related Works}
\label{sec:relatedwork}
There are quite a few studies on selective search on document sharding and query routing. Most of them aggregate similar documents into the same shards by measuring the semantic relevance among them in the collection to facilitate the subsequent search by limiting the query to be executed only within a few shards \citep{aly_taily_2013, kulkarni_efficient_2013, kulkarni_document_2010, kulkarni_selective_2015}. Only a few studies on document sharding utilize the search log, where the information within the queries used to retrieve the documents are tailored to assist the clustering \citep{puppin_query_2006, poblete_query_2008, dai_query_2016}. But they either merely extract simple features from the query to form rules or just employ the term frequency information in the queries as weights in the documents clustering process, neglecting both the strong connection between queries and documents and the semantic information hidden inside the query. Besides, all the previous studies rely on a multi-stage process for document clustering and the subsequent query routing, where an additional shard selection procedure has to be applied during the query routing stage \citep{si_relevant_2003, thomas_sushi_2009, kulkarni_shard_2012, aly_taily_2013}. Besides, earlier studies primarily focus on ensuring semantically similar documents are allocated in the same cluster, neglecting the imbalance among the generated clusters, which weakens their application in practice. Furthermore, only a few studies take load-balancing into account when allocating documents \citep{kim_load_2016, dai_query_2016}. But these load balancing algorithms are rule-based procedures (by first ordering the shards by sizes and then regrouping them), in addition to the document sharding step.

Co-training \citep{blum_combining_1998, Ganchev_2008_multi} was initially proposed for semi-supervised training, where two classifiers are trained separately on two distinct views of the same input data and forced to have similar predictions. These two classifiers are designed to constrain each other to make coherent decisions, and the objective function penalizes the disagreement between them. \citet{dasgupta_pac_2001} established the PAC generalization bounds for co-training for multiple classes classification, extending the theoretical bound on only two classes in the original study \citep{blum_combining_1998}. However, as pointed out by  \citet{pierce_limitations_2001}, the success of co-training heavily relies on the high-quality labeled data. This is also true in an attempt to apply co-training for unsupervised learning \citep{collins_unsupervised_1999}, where sophisticated rules which are manually extracted are used as seeds for bootstrapping.

Information Maximization (IM) \citep{gomes_discriminative_2010, bridle_unsupervised_1991}, promoting the idea that the output shall retain as much information as the input variable, has shed light on learning without direct supervision signals. In language processing, the celebrated brown clustering \citep{brown_class_1992} maximizes the mutual information between random bigrams. The information bottleneck approach \citep{tishby_information_1999} proposes an algorithm that discovers the representative coding of the input by capturing its relevant structure and proves its convergence. \citet{gomes_discriminative_2010} observe the degeneration and class imbalance when using mutual information for clustering and propose adding a regularization term for remedy. Combined with the rejuvenated deep neural networks, IM has shown robust performance on clustering \citep{hu_learning_2017} with data augmentation. 

More recent works have studied the application of information maximization in relational data. Co-training has been jointly used with information maximization in speech processing \citep{mcallester_information_2018}. \citet{stratos_mutual_2019} models the contextual information and the target class via mutual information maximization for word class induction, showing that the variational lower bound is more robust against the bias from noise. It is built upon the theoretical findings that the mutual information estimation through any high-probability lower bound has limitations that measuring and maximizing mutual information from finite data is a challenging training objective \citep{mcallester_formal_2018}. This study is further extended in an adversarial training setup to circumvent computing the global marginal distribution \citep{stratos_learning_2020}. Our work on selective search is inspired by these studies, treating the query and the document as two views of one sample and respecting the resulting cluster size balance. It is trained in an end-to-end paradigm without direct supervision signals.

In the IR community, whether dense or sparse retrieval is better is still a prolonged debate \citep{lin_pyserini_2021, luan_sparse_2021}, and recent studies have shown the improved performance of interpolating both \citep{li2022interpolate}, implying designing an appropriate approach to combine them is a promising direction. We would like to note selective search is the strategy for distributing the information retrieval workload, and is complementary and agnostic to innovations in algorithms related to dense or sparse retrieval. In fact, our selective search approach can be used in conjunction with dense, sparse or combined retrieval. 


\section{MICO: Mutual Information CO-training}
\label{sec:method}

Our assumption is that for a relevant query-document pair in the search, the cluster index assignments $z$ of the query $\bm{q}$ and $z'$ of the document $\bm{d}$ are two views of that pair $(\bm{q}, \bm{d})$, thus $z$ is equal to $z'$. MICO, consisting of a document allocation module and a query routing module, is trained based on this assumption. After training, MICO assigns the document to the cluster based on the prediction made by the document allocation module and routes new incoming queries to the cluster based on the prediction by the query routing module.


Our goal is to learn a model from the double-view examples in the search logs. The search logs are historical data that contains the relevance information between the query and its relevant documents, but does not carry explicit information of how documents shall be allocated into different shards, neither the query routing information of which shard the query shall be dispatched to. Thus, we have minimal supervision signal when training the model to learn to cluster documents and queries. 

\begin{figure}[h]
\centering
\includegraphics[width=0.50\textwidth, clip]{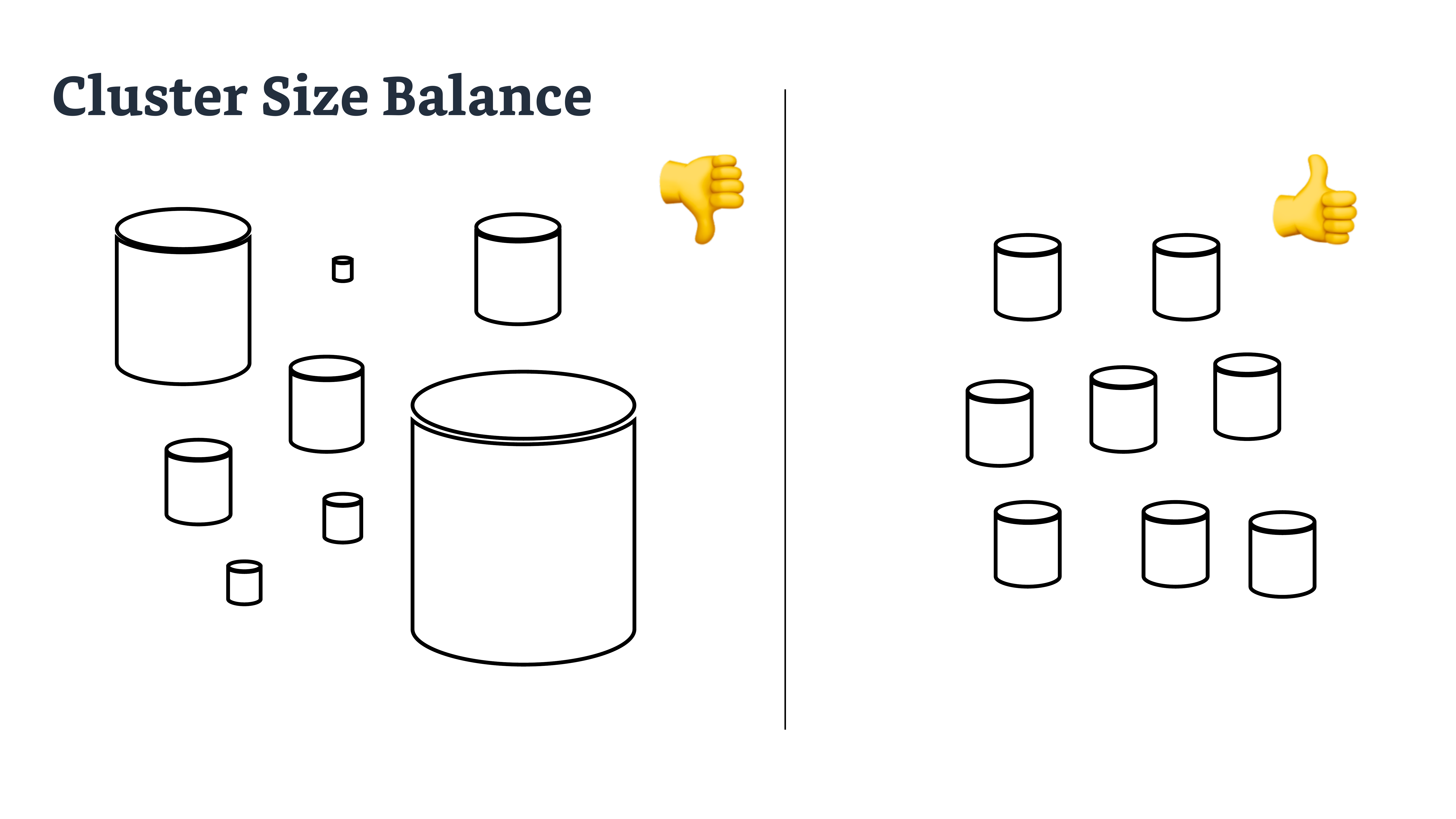}
\caption{We use entropy regularization to ensure the desired cluster size balance, thus document shard size balance is achieved by maximizing the entropy $H(Z')$.}
\label{fig:cluster_size_balance}
\end{figure}

The desired model maximizes the document coverage (recall) and minimizes the search cost simultaneously. We force the cluster selection of the query and the relevant document to be coherent during training, formally as $p(z|\bm{q}) \approx p(z'|\bm{d})$. Here we force the two distributions to be consistent if the query and the document are relevant to reflect the coherent cluster assignment. Meanwhile, to avoid certain resulting clusters being oversized, causing the search in these shards to be inefficient, we consider balancing the document shard sizes. This concept is illustrated in Figure \ref{fig:cluster_size_balance}. This balance is achieved by assigning the documents to clusters as uniform as possible (formally $p(z') \approx\mathrm{Uniform}$ where $p(z')$ is the distribution of cluster indices ${z}'$). MICO achieves cluster selection coherency between queries and documents by minimizing the mutual information between $p(z|\bm{q})$ and $p(z'|\bm{d})$, and balances shard size by adding an entropy regularization term $p(z')$.

After training, the model allocates all the documents into shards; and for a new coming query, since we have cast query routing as a clustering problem, the query routing module generates a score for an incoming query during the inference time across all the shards, and will route the query to the shards that have the highest scores.

MICO consists of three main components: One component maps queries to clusters, denoted by the conditional distribution $p^\psi(z|\bm{q})$, parameterized by $\psi$; Another one projects documents to clusters, denoted by the conditional distribution $p^\phi(z'|\bm{d})$, parameterized by $\phi$; The third one, denoted by $g^\theta(z')$ parameterized by $\theta$, uses a variational distribution to approximate an intractable distribution $p^\phi(z')$. All these three components are modeled via neural networks to account for the complexity.


\subsection{Tractable Loss of MICO}
{As the mutual information $I(Z;Z')$ of the query and the document cluster assignment and the entropy regularization $H(Z')$ of the document assignment are intractable (details are in the Appendix Section \ref{sec:info_max_approx}), we approximate them using $H(Z; Z')$ and $H^+(Z')$ respectively, i.e.,} 
we define the loss function of MICO as 
\begin{equation}
\mathcal{L} = {H}(Z';Z) - \beta{H}^+(Z')
\label{eq:mico_loss}
\end{equation}
where 
\begin{align*}
    H(Z';Z) = \mathbb{E}_{(\bm{q},\bm{d})\sim p_{QD}}[H(p^\phi(z'|\bm{d});p^\psi(z|\bm{q}))]\\
= \mathbb{E}_{(\bm{q},\bm{d})\sim p_{Q,D}} \left[\sum_{z=z'}- p^{\phi}(z'|\bm{d}) \log \left( p^{\psi}(z|\bm{q})\right)\right], \text{and}\\
{H}^+(Z') = \sum_{z'}-\mathbb{E}_{\bm{d}\sim p_{D}}[p^\phi(z'|\bm{d})] \log\left(g^\theta(z')\right), 
\end{align*}
with $\beta$ as a tunable hyper-parameter. 

The optimization objective is to find 
\begin{equation}
\phi^*, \psi^* = \argmin_{\phi,\psi} \left\{\max_\theta \mathcal{L}\right\}.
\end{equation}
We solve this minimax optimization problem by interleaving updating $(\phi,\psi)$ and $\theta$ during training. This is similar to training generative adversarial networks (GAN) in deep learning or the Actor-Critic model in reinforcement learning.

{In the rest of this section, we provide brief explanations of the loss above.}
We rewrite $\mathcal{L}$ as $-\left({I}^+(Z;Z') + (\beta-1){H}^+(Z')\right)$ where ${I}^+(Z;Z')=-{H}(Z';Z) + {H}^+(Z')$, ${I}^+(Z;Z')$ and ${H}^+(Z')$ are approximated upper bounds of ${I}(Z;Z')$ and ${H}(Z')$ respectively. {The intractable term ${H}(Z')$ is approximated by ${H}^+(Z')$, by replacing $\log(p^\phi(z'))$ with $\log(g^\theta(z'))$ ($\log(p^\phi(z'))$ is intractable due to the nature it is evaluated on full dataset): we maximize $H^+(Z')$ w.r.t. $\theta$ to force $g^\theta(z')$ to be close to $p^\phi(z')$.} Hence, minimizing $\mathcal{L}$ is equivalent to maximizing ${I}^+(Z;Z')$ and ${H}^+(Z')$ when $\beta > 1$. {In contrast to the proposed loss, direct application of the cross-entropy $H(Z';Z)$, corresponding to $\beta=0$, is inappropriate here, as it will result in cluster imbalance: the model tends to assign all the documents into one cluster and route all queries to this shard.}

The choice of $\beta$ in the loss $\mathcal{L}$ in Equation \ref{eq:mico_loss} controls the trade-off between the mutual information $I^+(Z;Z')$ and the entropy ${H}^+(Z')$, which in turn balances shard sizes while retaining the success of selective search. ${H}(Z')$ is maximized if and only if when $\mathbb{E}_{\bm{d}\sim p_{D}}[p^\phi(z'|\bm{d})]$ is a uniform distribution over $Z'$, i.e., the cluster sizes are all equivalent. Larger $\beta$ helps regularize our document allocation module to better balance the cluster sizes which leads to reduced search costs and latency.

\subsection{MICO-q: MICO with Query Consistency}
Multiple relevant documents can correspond to the same single query in the search log. We hypothesize that forcing the coherency of cluster assignment among these documents shall help the model yield a better document allocation module. Therefore, we use a cross-entropy $H^q$ among those documents as an additional regularization term. With this intuition, we develop a variant of MICO termed MICO-q, whose loss function is
\begin{align*}
\mathcal{L}^q &= - \beta{H}^+(Z') + \frac{1}{1+\gamma}({H}(Z';Z) + \gamma H^q), \text{with}\\
H^q & = \underset{\substack{\bm{q}\sim p_Q, \bm{d}_1,\bm{d}_2 \sim p_{D|Q}}}{\bbE} H\left(p^\phi(z|\bm{d}_1);p^\phi(z|\bm{d}_2)\right),
\end{align*}
where $\gamma$ is another tunable hyper-parameter.

\section{Experiments}
\label{sec:exps}
\subsection{Data sets}
We evaluate the proposed framework on the following two data sets, representing two different domains: the E-commerce search logs (ECSL) data set and the Cross-Lingual Information Retrieval (CLIR) data set derived from Wikipedia \citep{sasaki-etal-2018-cross}. We process the data to ensure the data sets contain only distinct documents and divide the queries as non-overlapped training, development, and testing set, following standard IR research procedures. We show the detailed data statistics in Table \ref{table:datastat} in Appendix.

\paragraph{ECSL} E-commerce search logs (ECSL) data set is created by sampling queries and product description documents on an English commercial shopping website. The
relevance of a document to a query is derived from user actions on the search results. We consider three types of user actions on documents with increasing levels of relevance to the query:
\textit{impression} (the document is among the search results), \textit{click} (the document link in the search results is clicked), and \textit{purchase} (the product of the document from the search result is purchased). The former relation always entails the latter.
We randomly split the queries into training, validation, and testing sets with the ratio of 8:1:1.

\paragraph{CLIR} In the CLIR set, queries are in English, extracted as
the first sentences from English wiki pages, with the title words removed. Two different types of relevance exist: \textit{DL} stands for directly relevant, meaning the documents are
the foreign-language pages having an inter-language link to the English pages; \textit{PL} stands for partially relevant, denoting the documents having mutual links to and from the \textit{DL} document. To provide a comprehensive study on the robustness of the proposed framework, we select the documents in two high-resource languages: French (\textit{fr}), Italian (\textit{it}) and in two low-resource languages: Tagalog (\textit{ta}), Swahili (\textit{sw}). Queries are randomly split into training, validation, and testing sets with the ratio of 3:1:1. 

\subsection{Experimental Setup}
In this study, we preset the number of shards to be 64 for the ECSL data set and 10 for the CLIR data set, according to the number of documents. To make a fair comparison with other baselines, we use the TF-IDF feature with the most frequent words as the feature vector for both the queries and the documents.

We compare our proposed methods with a number of competitive baselines in selective search, including \textbf{Random} assignment of documents into clusters, \textbf{K-means}, a variant of K-means called \textbf{Balanced K-means} ensuring balanced cluster sizes, an information-maximization based clustering algorithm \textbf{IMSAT} \citep{hu_learning_2017}, \textbf{KLD} sharding algorithm \citep{kulkarni_selective_2015} and \textbf{QKLD} sharding algorithm \citep{dai_query_2016}. Details of these baselines are in Appendix \ref{app:baselines}.

In our experiments, we use the \textit{impression} relationship between the training query and the documents for training in ECSL data. We use both the \textit{DL} and \textit{PL} relations for training in CLIR, as they are non-overlapped documents.
The expectation over the whole training set is approximated by the average over one batch during the training, t. We also provide a detailed theoretical analysis of batch training in MICO in Appendix \ref{app:sgd_mico}. Due to limited spaces, more details of the experimental setup can be found in Appendix \ref{app:experimental_setup}. Note neither the queries nor the documents have cluster labels, and the clustering effect in selective search is evaluated by the retrieval performance explained next.


\subsection{Query Coverage Analysis}
We first evaluate the query coverage (``recall'' in some literature) of our methods and compare it with baselines. Query coverage measures the percentage of documents retrieved in the top few shards over all the documents retrieved in the exhaustive search. 

Formally, the query coverage of a single query $q$ in the top $N$ shards is defined as 
$Cov_{N}(q) = \left(\sum_{i=1}^{N}R_{s_{i}^{q}}^{q}\right)/{R^{q}},$
where $R^{q}$ is the total number of relevant documents in the corpus per the query $q$, and $R_{s_{i}^{q}}^{q}$ is the number of relevant document per the query $q$ in the $i$-th selected shard $s^{q}_{i}$ per query $q$. Consequently, the average coverage over the query set is 
$Cov_{N} = \left(\sum_{q \in Q} Cov_{N}(q)\right)/{|Q|},$ 
where $Q$ is the testing queries set.

Our implemented coverage metric is even more strict than the coverage metric in \citet{dai_query_2016}'s study. Instead of sorting the shards by the number of retrieved documents they contain, we rigorously sort the chosen shards of a testing query based on the prediction score generated by the query routing module, which is more intrinsic to the nature of selective search. 

We show our experimental results on ECSL in Table \ref{table:query_coverage_on_ECSL}, with the mean value and the standard deviation in the parenthesis over five runs on three different relevances. We limit each testing query being routed to the most relevant \textbf{one} shard and \textbf{ten} shards based on the prediction. Our MICO and its variant MICO-q beat all the baselines, except \textit{impression} with one shard only. By only searching within the ten most relevant shards, which is 15.6\% of all the shards, MICO-q can achieve almost 95\% coverage on \textit{impression}, even comparable to the exhaustive search. MICO-q performs slightly more robust than MICO when searching over more shards, which might be attributed to its additional regularization term. 

We also present the evaluation results on CLIR in Table \ref{table:query_coverage_on_CLIR}, where we only probe the most relevant shard for each query. 
Due to the queries and documents being of different languages, we build vocabularies for the queries and documents separately, denoted by \textit{sv}, to generate input to the document sharding module and the query routing module. MICO (and MICO-q) beats all the baselines (except \textit{DL} in \textit{ta}). We also notice the performance of QKLD dramatically drops on the CLIR data set. We attribute this failure to its usage of the query information to build document shards inappropriately because queries and documents are from different languages. QKLD and MICO (including MICO-q) are the only two models to utilize the query information specifically for document sharding. Based on the observation, we argue inappropriate use of the query information will deteriorate document sharding in selectively search. MICO, on the contrary, is able to leverage the queries for better document sharding, showing its robustness in the cross-lingual scenario.

Though MICO beats all the baselines, if the top selected shards are extremely large, this approach is inapplicable in practice, as the actual search cost will be similar to the exhaustive search. In addition, MICO fails to beat KLD and QKLD in some exceptions, potentially ascribed to the most relevant shard created by them being larger with more documents. Thus, we analyze the search cost next.

\begin{table*}[!ht]
\centering
{\footnotesize
\begin{tabular}{l>{\centering\arraybackslash}p{6em}>{\centering\arraybackslash}p{6em}>{\centering\arraybackslash}p{6em}>{\centering\arraybackslash}p{6em}>{\centering\arraybackslash}p{6em}>{\centering\arraybackslash}p{6em}}
\toprule
  & \multicolumn{2}{c}{\makecell[c]{\textit{impression}}} & \multicolumn{2}{c}{\makecell[c]{\textit{click}}} & \multicolumn{2}{c}{\makecell[c]{\textit{purchase}}} \\
\cmidrule{2-7}
 Models & N=1 & N=10 & N=1 & N=10 & N=1 & N=10 \\
 \midrule
 Random & 1.56 (6e-3) & 15.62 (0.02) & 1.49 (0.08) & 15.32 (0.85) & 1.45 (0.24) & 14.54 (0.27) \\ 
  \midrule
 {K-means} & 48.98 (1.60) & 79.05 (0.51) & 51.90 (1.56) & 81.57 (4.0) & 54.49 (1.97) & 83.58 (1.49) \\
  \midrule
 \makecell[lt]{B-K-means} & 39.72 (1.12) & 64.56 (1.30) & 43.89 (2.03) & 64.25 (1.78) & 49.02 (2.37) & 69.59 (1.22) \\
  \midrule
 \makecell[lt]{IMSAT} & 41.68 (0.55) & 71.37 (0.28) & 47.48 (1.62) & 79.12 (2.94) & 52.41 (0.42) & 79.83 (1.06) \\
  \midrule
 KLD & 43.46 (5.91) & 69.87 (5.34) & 44.94 (8.04) & 71.17 (5.55) & 46.77 (9.32) & 70.5 (4.08) \\ 
\midrule
 QKLD & \textbf{86.14} (8.85) & 93.96 (0.77) & 73.72 (7.25) & 81.89 (1.2) & 75.79 (7.22) & 83.56 (1.57) \\ 
\midrule
\midrule
 MICO & 67.09 (0.20) & 92.85 (0.12) & \textbf{82.85} (1.51) & 97.81 (0.19) & \textbf{81.21} (0.49) & 96.61 (0.14) \\
   \midrule
 MICO-q & 69.81 (0.34) & \textbf{94.28} (0.09) & 82.48 (1.91) & \textbf{98.26} (0.20) & 81.15 (1.23) & \textbf{97.25} (0.16) \\
 \bottomrule
\end{tabular}
}
\caption{This table shows the performance of query coverage (recall) of MICO, MICO-q, and different baselines over three different query-document relationships on the ECSL data set. We show the performance by only probing the top-1 most relevant shard and the top-10 most relevant shards given a query. The number in the parenthesis right next to the coverage is the standard deviation over five runs. We observe other than the impression relation in which QKLD has the best performance, MICO or MICO-q beat all the baselines.}
\label{table:query_coverage_on_ECSL}
\end{table*}

\begin{table*}[!ht]
\centering
{\scriptsize
\begin{tabular}{l>{\centering\arraybackslash}p{5em}>{\centering\arraybackslash}p{5em}>{\centering\arraybackslash}p{5em}>{\centering\arraybackslash}p{5em}>{\centering\arraybackslash}p{5em}>{\centering\arraybackslash}p{5em}>{\centering\arraybackslash}p{5em}>{\centering\arraybackslash}p{5em}}
\toprule
  & \multicolumn{2}{c}{\makecell[c]{\textit{fr}}} & \multicolumn{2}{c}{\makecell[c]{\textit{it}}} & \multicolumn{2}{c}{\makecell[c]{\textit{ta}}} & \multicolumn{2}{c}{\makecell[c]{\textit{sw}}} \\
\cmidrule{2-9}
 Models & \textit{DL} & \textit{PL} & \textit{DL} & \textit{PL} & \textit{DL} & \textit{PL} & \textit{DL} & \textit{PL} \\
 \midrule
 Random & 10.02 (0.07) & 9.72 (0.16) & 10.02 (0.09) & 10.0 (0.35) & 9.88 (0.93) & 9.86 (0.48) & 10.01 (0.23) & 10.0 (0.67) \\ 
  \midrule
 {K-means} & 12.19 (1.99) & 10.79 (2.04) & 14.91 (2.46) & 16.36 (3.55) & 16.25 (2.5) & 21.08 (3.46) & 21.71 (4.84) & 18.55 (3.65) \\
  \midrule
 \makecell[lt]{B-K-means} & 12.2 (1.82) & 11.44 (1.32) & 12.45 (3.13) & 12.46 (4.59) & 12.78 (2.85) & 11.23 (3.84) & 11.71 (1.29) & 12.16 (1.21) \\
  \midrule
 \makecell[lt]{IMSAT} & 19.77 (9.53) & 19.84 (9.68) & 40.09 (8.91) & 40.13 (8.88) & 12.72 (4.22) & 11.89 (3.62) & 8.4 (2.24) & 8.6 (2.63)\\
  \midrule
 KLD & 38.6 (6.02) & 40.65 (7.58) & 60.94 (5.25) & 61.83 (3.12) & \textbf{66.53} (8.43) & 59.77 (7.18) & 21.11 (3.52) & 24.83 (3.81)\\ 
\midrule
 QKLD & 17.76 (3.63) & 18.82 (2.08) & 18.9 (4.91) & 17.45 (4.12) & 23.65 (6.81) & 24.4 (5.54) & 12.23 (0.75) & 16.45 (2.25) \\ 
  \midrule
  \midrule
 MICO (sv) & 44.93 (3.47) & \textbf{53.12} (2.17) & 58.08 (1.22) & 65.83 (1.06) & 63.55 (4.45) & 60.94 (4.92) & 26.0 (3.51) & \textbf{28.67} (3.73) \\
  \midrule
 MICO-q (sv) & \textbf{47.9} (2.68) & 48.04 (3.44) & \textbf{75.27} (3.6) & \textbf{75.01} (4.39) & 63.91 (5.3) & \textbf{61.29} (5.31) & \textbf{27.42} (3.37) & 28.14 (2.54) \\
 \bottomrule
\end{tabular}
}
\caption{This table shows the performance of query coverage of MICO, MICO-q, and different baselines on two different query-document relationships on the CLIR data set by only probing the most relevant shard given a query because we only divide the documents into ten shards. The number in the parenthesis right next to the coverage is the standard deviation over multiple runs. \textit{sv} stands for separate vocabularies for the queries and documents, as in cross-lingual retrieval, the source language and the target language have different vocabularies, and separate vocabularies perform better than unified ones empirically. MICO and MICO-q beat all the baselines except \textit{DL} in \textit{ta}.}
\label{table:query_coverage_on_CLIR}
\end{table*}

\subsection{Cost Analysis}

To measure the search efficiency, we use the metrics \textbf{search resource cost} and \textbf{search latency cost} (the lower, the better) introduced in the early study \citep{kulkarni_shard_2012} to evaluate our models and the competitors. Search cost depends highly on the size of the shards we send the query to. Search resource cost calculates the resource usage as the upper bound on the number of document evaluated for each query, defined as $C^{res}_{N} = \sum_{i=1}^{N} |s_{i}^q|,$ where $|s_{i}^q|$ is the number of documents in the $i$-th relevant cluster to query $q$.
Similarly, search latency cost counts the number of evaluated documents on the longest execution path for query $q$ for its top $N$ selected shards: $C^{lat}_{N} = \max_{1\leq i \leq N} |s_{i}^q|.$\footnote{Note the metrics in \citet{kulkarni_shard_2012} introduce an additional term for the resource selection step. Since MICO doesn’t require this additional step, that term is always evaluated to 0 with MICO (and MICO-q).}

We show our evaluation results on the ECSL data set and the CLIR data set in Table \ref{table:seach_cost}, by restricting the search within the \textbf{five} most relevant clusters for ECSL, and the \textbf{two} most relevant for CLIR. Please note in these evaluations \textbf{Random} is the skyline, as all the documents are evenly distributed to $N$ shards such that the shard sizes are almost equal. MICO demonstrates its supreme performance, even beating the skyline in some cases, showing its ability to control shard sizes while ensuring semantically similar documents are allocated in the same shard. MICO-q also achieves comparable performance with MICO. We interpret it as MICO and MICO-q are able to trade off document relevance in the same shard and balance of shard sizes on a sweet point. We also notice the high cost yielded by KLD and QKLD, which compromises their use in practice. 

\begin{table*}[!ht]
\centering
{\footnotesize
\begin{tabular}{l>{\centering\arraybackslash}p{3em}>{\centering\arraybackslash}p{3em}>{\centering\arraybackslash}p{3em}>{\centering\arraybackslash}p{3em}>{\centering\arraybackslash}p{3em}>{\centering\arraybackslash}p{3em}>{\centering\arraybackslash}p{3em}>{\centering\arraybackslash}p{3em}>{\centering\arraybackslash}p{3em}>{\centering\arraybackslash}p{3em}}
\toprule
\multirow{2}{*}{} & \multicolumn{5}{c}{$C^{res}_{N}$} & \multicolumn{5}{c}{$C^{lat}_{N}$}\\
\cmidrule{2-5}
\cmidrule{6-11}
 Models & ECSL & C-fr & C-it & C-ta & C-sw & ECSL & C-fr & C-it & C-ta & C-sw \\
  \midrule
 K-means  & 2.061 & 14.12 & 11.54 & 1.6 & 1.42 & 1.572 & 6.44 & 5.95 & 0.95 & 1.27 \\
  \midrule
 Balaced K-means  & 0.620 & 8.1 & 6.83 & 0.99 & 0.87 & 0.277 & 2.58 & 2.02 & 0.34 & 0.67 \\
  \midrule
 IMSAT  & 0.370 & 9.57 & 5.89 & 0.93 & 0.61 & 0.082 & 3.57 & 4.78 & 0.58 & 0.53 \\
  \midrule
 KLD & 2.17 & 17.48 & 13.15 & 1.93 & 1.09 & 1.41 & 13.26 & 11.43 & 1.72 & 0.74 \\ 
  \midrule
 QKLD & 4.5 & 8.84 & 7.42 & 1.34 & 0.99 & 4.47 & 3.72 & 3.07 & 0.94 & 0.66 \\ 
  \midrule
  \midrule
 MICO & \textbf{0.367} & \textbf{6.19} & \textbf{5.13} & 0.85 & 0.93 & 0.089 & \textbf{2.34} & \textbf{1.89} & 0.5 & 0.5 \\
  \midrule
 MICO-q & 0.369 & 7.12 & 6.71 & 0.94 & 1.07 & 0.093 & 2.73 & 2.47 & 0.51 & 0.58 \\
  \midrule
  \midrule
 Random & 0.368 & 7.20 & 5.95 & \textbf{0.8} & \textbf{0.73} & \textbf{0.074} & 2.41 & 1.99 & \textbf{0.27} & \textbf{0.25} \\ 
 \bottomrule
\end{tabular}
}
\caption{This table shows the performance of different models on the \textit{Search Resource Cost} and the \textit{Search Latency Cost} metrics, representing the search efficiency, with the lower the number, the better the performance. The results shown in this table are scaled by being divided by $10^6$ on the ECSL data set and by $10^4$ on the CLIR data set. Note in this set of experiments, we use separate vocabulary (\textit{sv}) for MICO and MICO-q on CLIR. We observe the supreme performance of MICO, which in some cases even beats the \textbf{Random} skyline.}
\label{table:seach_cost}
\end{table*}




\subsection{Balance Among Shard Sizes}
\begin{figure}[!ht]	
	\centering
	\begin{subfigure}[t]{0.22\textwidth}
		\centering
		\includegraphics[scale=0.25]{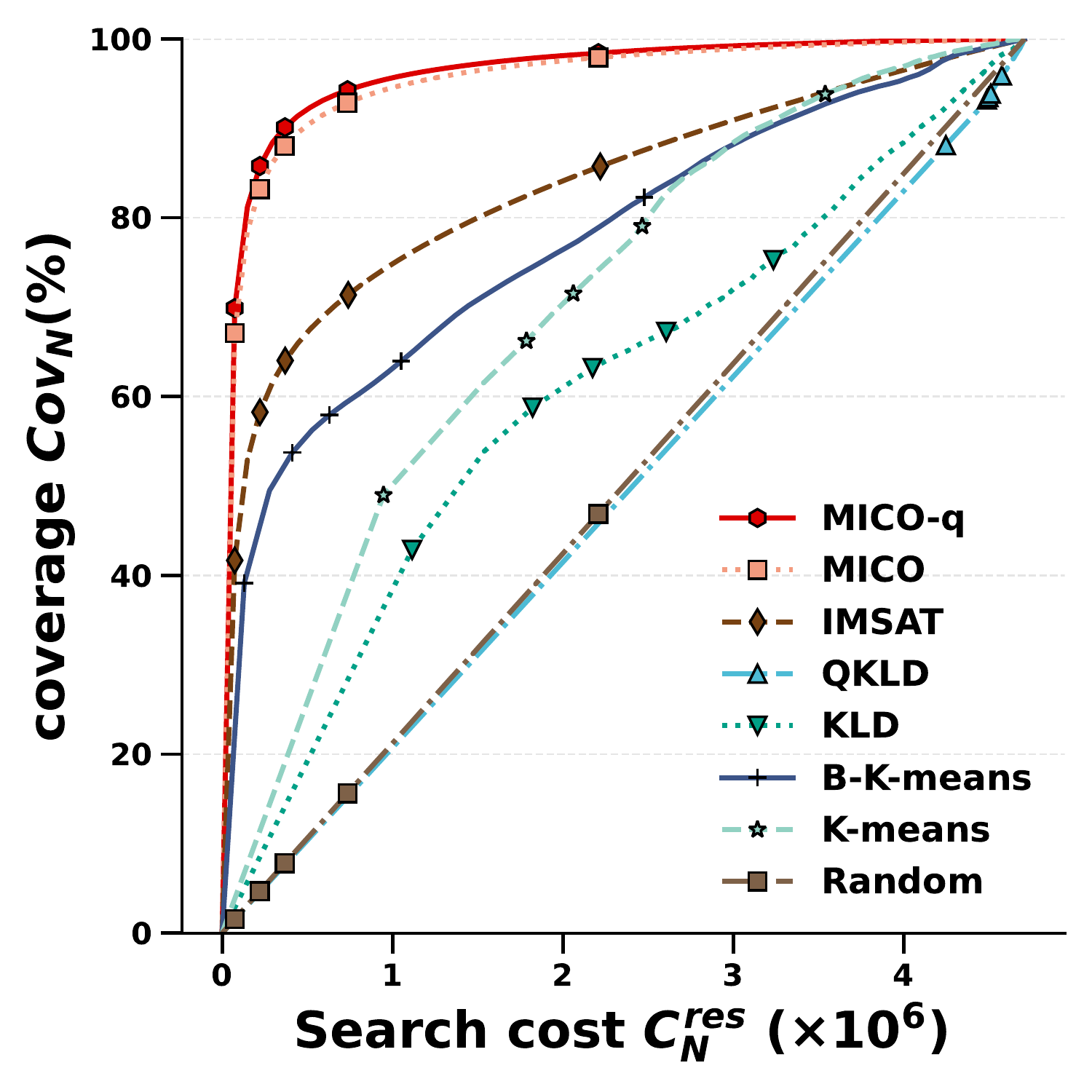}
		\caption{Coverage v.s. Cost}
		\label{fig:comparison_among_all_methods_a}
	\end{subfigure}
	\quad
	\begin{subfigure}[t]{0.22\textwidth}
		\centering
	    \includegraphics[scale=0.42]{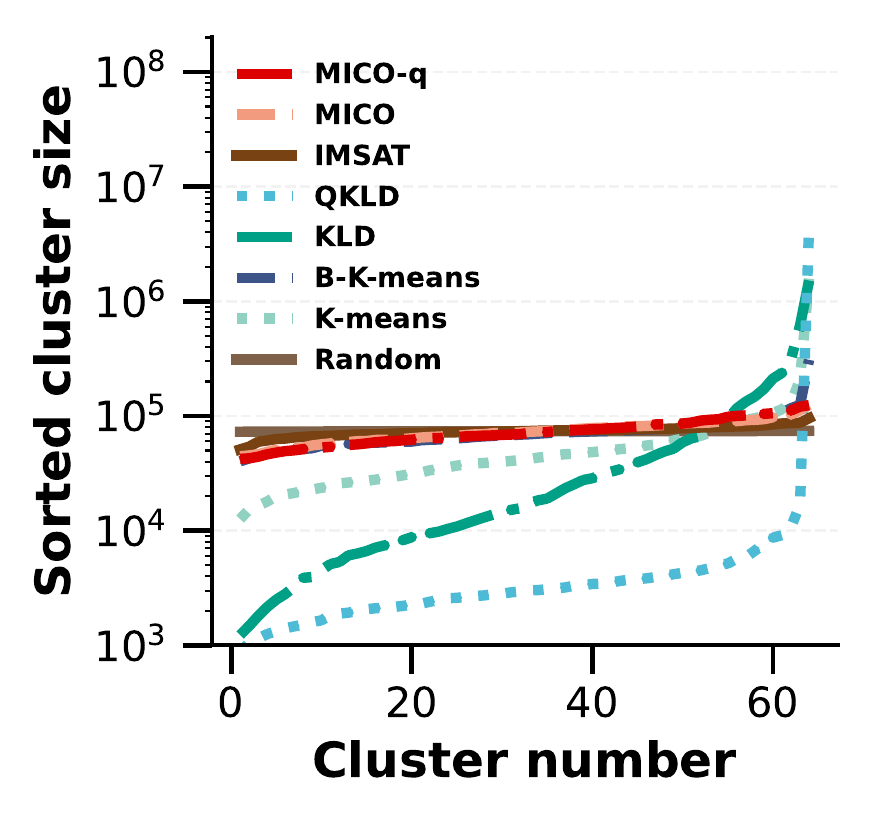}
		\caption{Shard size distribution}
		\label{fig:comparison_among_all_methods_b}
	\end{subfigure}
 \caption{Figure \ref{fig:comparison_among_all_methods_a} shows MICO and MICO-q are significantly better than all other methods as they have high \textit{impression} coverage with low search cost. From bottom-left to top-right, the markers on each line represent query coverage limited within the top-1, top-3, top-5, top-10, and top-30 clusters selectively. Figure \ref{fig:comparison_among_all_methods_b} shows \textbf{Random} generates the most balanced shard sizes (as a flat line), and IMSAT also creates very balanced shards. MICO and MICO-q are on a par with IMSAT. In contrast, QKLD and KLD yield very unbalanced shards.}
 \label{fig:comparison_among_all_methods}
\end{figure}

We further investigate the performance of MICO and MICO-q via visualizing the trade-off between \textit{query coverage} and \textit{search resource cost} and the balance of resulting shard sizes on ECLS. Figure \ref{fig:comparison_among_all_methods_a} shows that MICO and MICO-q always perform the best in terms of query coverage, at different levels of search resource cost. Figure \ref{fig:comparison_among_all_methods_b} indicates that IMSAT is able to create well balanced shards, and MICO and MICO-q are on a par with it. 
To trade off query coverage and shard size balance, MICO and MICO-q are the best among all the competitors, gaining the favor in practical usage. Note that KLD and QKLD create very unbalanced shards, which explains MICO and MICO-q fail to beat them in some cases on query coverage analysis earlier: most of the documents are allocated into the same shard.

\subsection{Effect of Entropy Regularization in MICO and MICO-q}
\begin{figure}[!htb]
	\centering
	\begin{subfigure}[t]{0.22\textwidth}
		\centering
		\includegraphics[width=\textwidth, ]{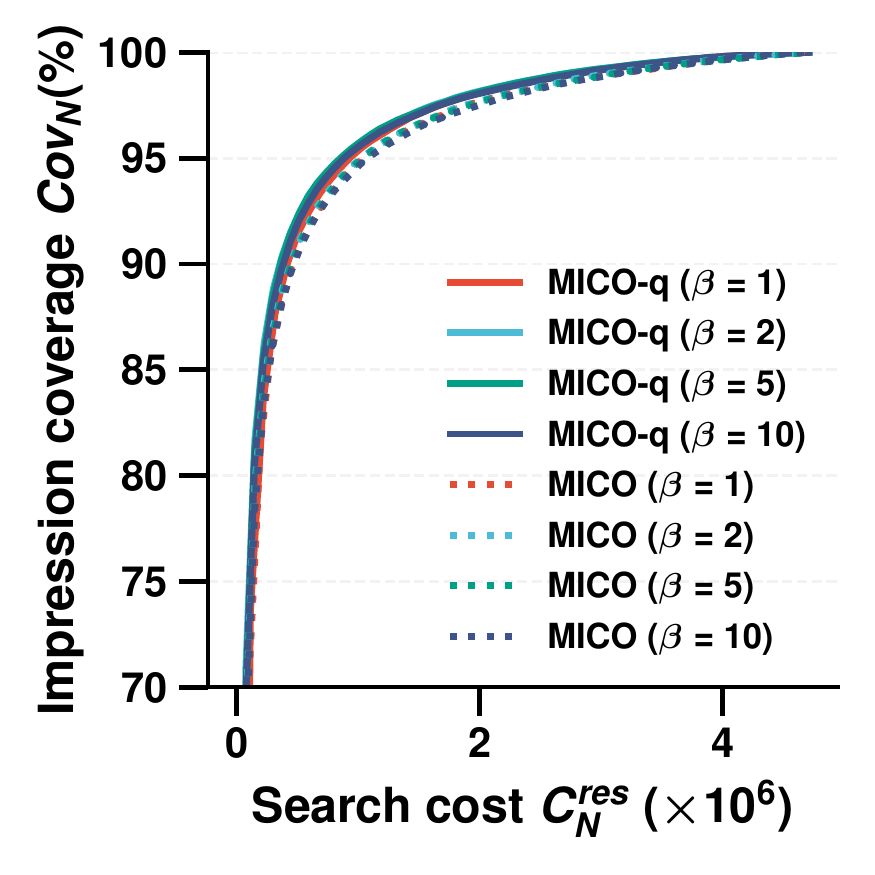}
		\caption{Coverage v.s. Cost}\label{fig:entropy_parameter_and_rebalance_a}		
	\end{subfigure}
	\quad
	\begin{subfigure}[t]{0.22\textwidth}
		\centering
	    \includegraphics[width=\textwidth,  ]{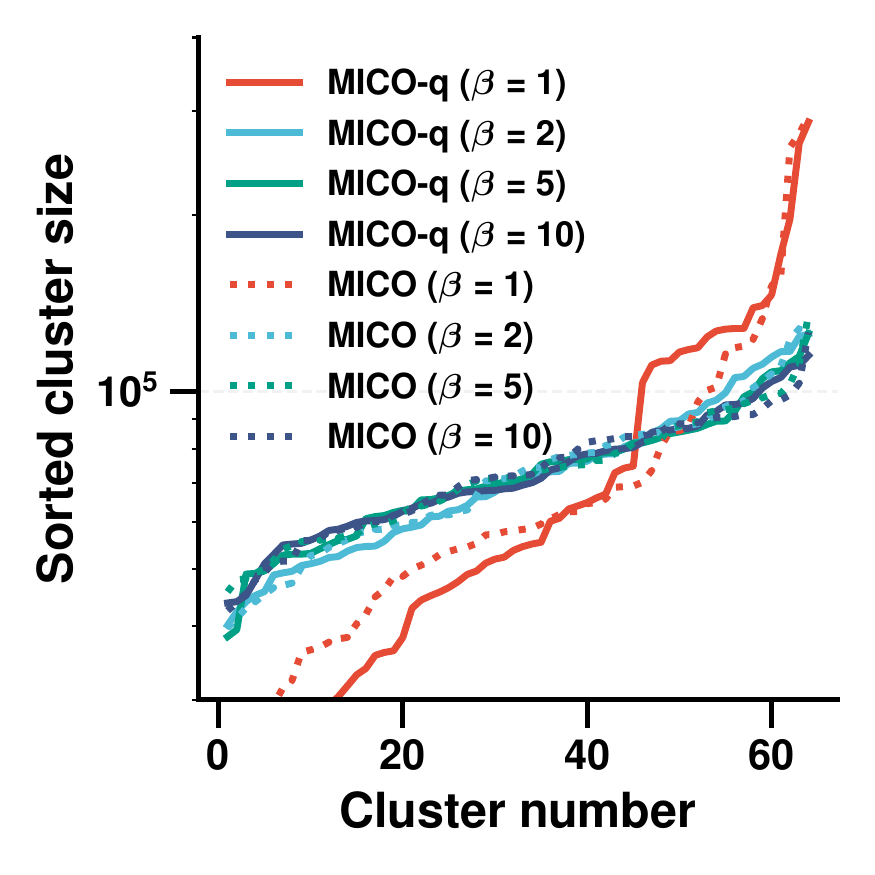}
		\caption{Shard size distribution}\label{fig:entropy_parameter_and_rebalance_b}		
	\end{subfigure}
 \caption{This figure shows the results of MICO and MICO-q with different entropy parameter $\beta\in\{1,2,5,10\}$. (a) The overall performance is hardly affected by $\beta$. (b) While larger $\beta$ induces better balance on the cluster sizes, the difference between $\beta=5$ and $\beta=10$ is very small. When setting $\beta=1$, the objective degenerates to use mutual information only (without the additional entropy regularization term) and the resulting shard sizes are significantly more unbalanced than with the other choices of $\beta$.
 }\label{fig:entropy_parameter_and_rebalance}
\end{figure}

We also examine the effect of entropy regularization by adjusting its strength in MICO and MICO-q by tuning their hyper-parameter $\beta$ on ECSL. From Figure \ref{fig:entropy_parameter_and_rebalance_a}, we can see that $\beta$ does not affect the cost-coverage curve in a significant way. From Figure \ref{fig:entropy_parameter_and_rebalance_b}, setting $\beta$ larger than 1 yields much more balanced cluster sizes which means that the entropy regularization is necessary to balance cluster sizes. We conjecture this difference results from the degree of homogeneity of the documents the model allocates to different shards.

\subsection{Ablation Study}
We further investigate the effectiveness of different neural architecture variants of MICO. Instead of parameterizing the document allocation module and the query routing module differently by two separate neural networks ($p_q^\psi(z|\bm{q}), p_d^\phi(z'|\bm{d})$), we can parameterize both modules with the same neural network ($p_q^\psi(z|\bm{q}), p_d^\psi(z'|\bm{d})$), avoiding the risk of over-parameterization. We denote this variant as \textbf{-Par}. The recent emergence of the pre-trained language models such as BERT \citep{devlin-etal-2019-bert} also provides a richer representation with contextual information. We use the BERT multilingual base model\footnote{https://huggingface.co/bert-base-multilingual-cased} to produce the text-level representations and feed them into MICO, either as fixed (denoted as \textbf{+BERT.fx}) or by fine-tuning the underlying BERT model (denoted as \textbf{+BERT.ft}). We show the query coverage (QC) and Search Resource Cost of these neural variants in Table \ref{table:ablation} on ECSL, by probing the most relevant shard for each query. The results show that unifying the query module and the document module parameterization deteriorates the performance. We also notice that fixed BERT representations hurt the model performance significantly, while appropriate fine-tuning help improve the performance.

\begin{table}[!ht]
\centering
\begin{tabular}{|l|p{3.5em}|p{3.5em}|}
\hline
 Model & QC & $C^{res}_{N=1}$ \\
  \hline
 MICO (-Par) & 66.08 & 0.365  \\ 
  \hline
 MICO & 67.09 & 0.367  \\ 
  \hline
  MICO (+BERT.fx) & 41.67 & 0.367  \\ 
  \hline
 MICO (+BERT.ft) & \textbf{76.41} & 0.375  \\ 
  \hline
\end{tabular}
\caption{This table shows MICO with neural architecture variants. BERT with fine-tuning achieves better performance than the original MICO, while the other variants yield deteriorated performance. The search cost is slightly higher with the best-performing system. We attribute that the refined representations cause the model to weigh more on semantic similarity than cluster balance.}
\label{table:ablation}
\end{table}



\section{Conclusion}
\label{sec: concl}
In this study, we present MICO, a Mutual Information CO-training framework for document sharding and query routing in selective search with minimal supervision. 
Our contributions are: First, during training, MICO maximizes the mutual information between the cluster assignment variables of the query and the document, forcing the prediction of two views of the query-document pair coherently. 
Second, this design enables it to be trained end-to-end for both document sharding (clustering) and subsequent query routing, featuring its practical use when vast volumes of documents are required to be searched simultaneously. 
Third, we show improved performance of MICO empirically on multiple important metrics in selective search. 

Our future research direction includes further investigation of the regularization for more balanced sizes among all the shards, the utilization of complex neural language models to enrich the document and the query representation for higher query coverage via better clustering results, the reduction of potential noise between a query and its associated document in the data set, and the detection policy to decide when the model needs to be retrained if constant update of the model is needed when deployed. We are also interested in potential extension of this approach to multi-modal data, e.g., image search, music search and software search where the non-traditional documents are in the non-textual formats.

\bibliography{mybib}

\begin{thebibliography}{34}
\expandafter\ifx\csname natexlab\endcsname\relax\def\natexlab#1{#1}\fi

\bibitem[{Aly et~al.(2013)Aly, Hiemstra, and Demeester}]{aly_taily_2013}
Robin Aly, Djoerd Hiemstra, and Thomas Demeester. 2013.
\newblock \href {https://doi.org/10.1145/2484028.2484033} {Taily: shard
  selection using the tail of score distributions}.
\newblock In \emph{The 36th International {ACM} {SIGIR} conference on research
  and development in Information Retrieval, {SIGIR} '13, Dublin, Ireland - July
  28 - August 01, 2013}, pages 673--682. {ACM}.

\bibitem[{Barroso et~al.(2003)Barroso, Dean, and H{\"o}lzle}]{barroso2003web}
Luiz~Andr{\'e} Barroso, Jeffrey Dean, and Urs H{\"o}lzle. 2003.
\newblock Web search for a planet: The google cluster architecture.
\newblock \emph{IEEE micro}, 23(2):22--28.

\bibitem[{{Blum} and {Mitchell}(1998)}]{blum_combining_1998}
Avrim {Blum} and Tom {Mitchell}. 1998.
\newblock Combining labeled and unlabeled data with co-training.
\newblock In \emph{Proceedings of the eleventh annual conference on
  Computational learning theory}, pages 92--100.

\bibitem[{Bridle et~al.(1992)Bridle, Heading, and
  MacKay}]{bridle_unsupervised_1991}
John Bridle, Anthony Heading, and David MacKay. 1992.
\newblock \href
  {https://proceedings.neurips.cc/paper/1991/file/a8abb4bb284b5b27aa7cb790dc20f80b-Paper.pdf}
  {Unsupervised classifiers, mutual information and \textquotesingle phantom
  targets}.
\newblock In \emph{Advances in Neural Information Processing Systems},
  volume~4. Morgan-Kaufmann.

\bibitem[{{Brown} et~al.(1992){Brown}, {deSouza}, {Mercer}, {Pietra}, and
  {Lai}}]{brown_class_1992}
Peter~F. {Brown}, Peter~V. {deSouza}, Robert~L. {Mercer}, Vincent J.~Della
  {Pietra}, and Jenifer~C. {Lai}. 1992.
\newblock Class-based n -gram models of natural language.
\newblock \emph{Computational Linguistics}, 18(4):467--479.

\bibitem[{Collins and Singer(1999)}]{collins_unsupervised_1999}
Michael Collins and Yoram Singer. 1999.
\newblock \href {https://www.aclweb.org/anthology/W99-0613} {Unsupervised
  {Models} for {Named} {Entity} {Classification}}.
\newblock In \emph{1999 {Joint} {SIGDAT} {Conference} on {Empirical} {Methods}
  in {Natural} {Language} {Processing} and {Very} {Large} {Corpora}}.

\bibitem[{Dai et~al.(2016)Dai, Xiong, and Callan}]{dai_query_2016}
Zhuyun Dai, Chenyan Xiong, and Jamie Callan. 2016.
\newblock \href {https://doi.org/10.1145/2983323.2983706} {Query-biased
  partitioning for selective search}.
\newblock In \emph{Proceedings of the 25th {ACM} International Conference on
  Information and Knowledge Management, {CIKM} 2016, Indianapolis, IN, USA,
  October 24-28, 2016}, pages 1119--1128. {ACM}.

\bibitem[{Dasgupta et~al.(2001)Dasgupta, Littman, and
  McAllester}]{dasgupta_pac_2001}
Sanjoy Dasgupta, Michael~L. Littman, and David~A. McAllester. 2001.
\newblock \href
  {https://proceedings.neurips.cc/paper/2001/hash/4c144c47ecba6f8318128703ca9e2601-Abstract.html}
  {{PAC} generalization bounds for co-training}.
\newblock In \emph{Advances in Neural Information Processing Systems 14 [Neural
  Information Processing Systems: Natural and Synthetic, {NIPS} 2001, December
  3-8, 2001, Vancouver, British Columbia, Canada]}, pages 375--382. {MIT}
  Press.

\bibitem[{Devlin et~al.(2019)Devlin, Chang, Lee, and
  Toutanova}]{devlin-etal-2019-bert}
Jacob Devlin, Ming-Wei Chang, Kenton Lee, and Kristina Toutanova. 2019.
\newblock \href {https://aclanthology.org/N19-1423} {{BERT}: Pre-training of
  deep bidirectional transformers for language understanding}.
\newblock In \emph{Proceedings of the 2019 Conference of the North {A}merican
  Chapter of the Association for Computational Linguistics: Human Language
  Technologies, Volume 1 (Long and Short Papers)}, pages 4171--4186,
  Minneapolis, Minnesota. Association for Computational Linguistics.

\bibitem[{Ganchev et~al.(2008)Ganchev, Gra\c{c}a, Blitzer, and
  Taskar}]{Ganchev_2008_multi}
Kuzman Ganchev, Jo\~{a}o~V. Gra\c{c}a, John Blitzer, and Ben Taskar. 2008.
\newblock Multi-view learning over structured and non-identical outputs.
\newblock In \emph{Proceedings of the Twenty-Fourth Conference on Uncertainty
  in Artificial Intelligence}, UAI'08, page 204–211, Arlington, Virginia,
  USA. AUAI Press.

\bibitem[{Gomes et~al.(2010)Gomes, Krause, and
  Perona}]{gomes_discriminative_2010}
Ryan Gomes, Andreas Krause, and Pietro Perona. 2010.
\newblock \href
  {https://proceedings.neurips.cc/paper/2010/hash/42998cf32d552343bc8e460416382dca-Abstract.html}
  {Discriminative clustering by regularized information maximization}.
\newblock In \emph{Advances in Neural Information Processing Systems 23: 24th
  Annual Conference on Neural Information Processing Systems 2010. Proceedings
  of a meeting held 6-9 December 2010, Vancouver, British Columbia, Canada},
  pages 775--783. Curran Associates, Inc.

\bibitem[{Gravano et~al.(1999)Gravano, Garc\'{\i}a-Molina, and
  Tomasic}]{gravano-1999-gloss}
Luis Gravano, H\'{e}ctor Garc\'{\i}a-Molina, and Anthony Tomasic. 1999.
\newblock \href {https://doi.org/10.1145/320248.320252} {Gloss: Text-source
  discovery over the internet}.
\newblock \emph{ACM Trans. Database Syst.}, 24(2):229–264.

\bibitem[{Hu et~al.(2017)Hu, Miyato, Tokui, Matsumoto, and
  Sugiyama}]{hu_learning_2017}
Weihua Hu, Takeru Miyato, Seiya Tokui, Eiichi Matsumoto, and Masashi Sugiyama.
  2017.
\newblock \href {http://proceedings.mlr.press/v70/hu17b.html} {Learning
  discrete representations via information maximizing self-augmented training}.
\newblock In \emph{Proceedings of the 34th International Conference on Machine
  Learning, {ICML} 2017, Sydney, NSW, Australia, 6-11 August 2017}, volume~70
  of \emph{Proceedings of Machine Learning Research}, pages 1558--1567. {PMLR}.

\bibitem[{Kim et~al.(2016)Kim, Callan, Culpepper, and Moffat}]{kim_load_2016}
Yubin Kim, Jamie Callan, J.~Shane Culpepper, and Alistair Moffat. 2016.
\newblock \href {https://doi.org/10.1145/2911451.2914689} {Load-balancing in
  distributed selective search}.
\newblock In \emph{Proceedings of the 39th International ACM SIGIR Conference
  on Research and Development in Information Retrieval}, SIGIR '16, page
  905–908, New York, NY, USA. Association for Computing Machinery.

\bibitem[{Kulkarni(2013)}]{kulkarni_efficient_2013}
Anagha Kulkarni. 2013.
\newblock \emph{Efficient and {Effective} {Large}-scale {Search}}.
\newblock {PhD} thesis, Carnegie Mellon University.

\bibitem[{Kulkarni and Callan(2010)}]{kulkarni_document_2010}
Anagha Kulkarni and Jamie Callan. 2010.
\newblock \href {https://doi.org/10.1145/1871437.1871497} {Document allocation
  policies for selective searching of distributed indexes}.
\newblock In \emph{Proceedings of the 19th ACM International Conference on
  Information and Knowledge Management}, CIKM '10, page 449–458, New York,
  NY, USA. Association for Computing Machinery.

\bibitem[{Kulkarni and Callan(2015)}]{kulkarni_selective_2015}
Anagha Kulkarni and Jamie Callan. 2015.
\newblock \href {https://doi.org/10.1145/2738035} {Selective {Search}:
  {Efficient} and {Effective} {Search} of {Large} {Textual} {Collections}}.
\newblock \emph{ACM Trans. Inf. Syst.}, 33(4).
\newblock Place: New York, NY, USA Publisher: Association for Computing
  Machinery.

\bibitem[{Kulkarni et~al.(2012)Kulkarni, Tigelaar, Hiemstra, and
  Callan}]{kulkarni_shard_2012}
Anagha Kulkarni, Almer~S. Tigelaar, Djoerd Hiemstra, and Jamie Callan. 2012.
\newblock \href {https://doi.org/10.1145/2396761.2396833} {Shard ranking and
  cutoff estimation for topically partitioned collections}.
\newblock In \emph{21st {ACM} International Conference on Information and
  Knowledge Management, CIKM'12, Maui, HI, USA, October 29 - November 02,
  2012}, pages 555--564. {ACM}.

\bibitem[{Li et~al.(2022)Li, Wang, Zhuang, Mourad, Ma, Lin, and
  Zuccon}]{li2022interpolate}
Hang Li, Shuai Wang, Shengyao Zhuang, Ahmed Mourad, Xueguang Ma, Jimmy Lin, and
  Guido Zuccon. 2022.
\newblock \href {https://doi.org/10.1145/3477495.3531884} {To interpolate or
  not to interpolate: Prf, dense and sparse retrievers}.
\newblock In \emph{Proceedings of the 45th International ACM SIGIR Conference
  on Research and Development in Information Retrieval}, SIGIR '22, page
  2495–2500, New York, NY, USA. Association for Computing Machinery.

\bibitem[{Lin(2021)}]{lin_pyserini_2021}
Jimmy Lin. 2021.
\newblock Pyserini: {A} {Python} {Toolkit} for {Reproducible} {Information}
  {Retrieval} {Research} with {Sparse} and {Dense} {Representations}.
\newblock In \emph{Proc. of {SIGIR}}, page~7.

\bibitem[{Luan et~al.(2021)Luan, Eisenstein, Toutanova, and
  Collins}]{luan_sparse_2021}
Yi~Luan, Jacob Eisenstein, Kristina Toutanova, and Michael Collins. 2021.
\newblock \href {https://doi.org/10.1162/tacl_a_00369} {Sparse, {Dense}, and
  {Attentional} {Representations} for {Text} {Retrieval}}.
\newblock \emph{Transactions of the Association for Computational Linguistics},
  9:329--345.

\bibitem[{McAllester(2018)}]{mcallester_information_2018}
David McAllester. 2018.
\newblock Information {Theoretic} {Co}-{Training}.
\newblock Eprint: 1802.07572.

\bibitem[{McAllester and Stratos(2020)}]{mcallester_formal_2018}
David McAllester and Karl Stratos. 2020.
\newblock \href {http://proceedings.mlr.press/v108/mcallester20a.html} {Formal
  limitations on the measurement of mutual information}.
\newblock In \emph{The 23rd International Conference on Artificial Intelligence
  and Statistics, {AISTATS} 2020, 26-28 August 2020, Online [Palermo, Sicily,
  Italy]}, volume 108 of \emph{Proceedings of Machine Learning Research}, pages
  875--884. {PMLR}.

\bibitem[{Pierce and Cardie(2001)}]{pierce_limitations_2001}
David Pierce and Claire Cardie. 2001.
\newblock \href {https://www.aclweb.org/anthology/W01-0501} {Limitations of
  co-training for natural language learning from large datasets}.
\newblock In \emph{Proceedings of the 2001 Conference on Empirical Methods in
  Natural Language Processing}.

\bibitem[{Poblete and Baeza-Yates(2008)}]{poblete_query_2008}
Barbara Poblete and Ricardo Baeza-Yates. 2008.
\newblock \href {https://doi.org/10.1145/1367497.1367504} {Query-sets: Using
  implicit feedback and query patterns to organize web documents}.
\newblock In \emph{Proceedings of the 17th International Conference on World
  Wide Web}, WWW '08, page 41–50, New York, NY, USA. Association for
  Computing Machinery.

\bibitem[{Puppin et~al.(2006)Puppin, Silvestri, and
  Laforenza}]{puppin_query_2006}
Diego Puppin, Fabrizio Silvestri, and Domenico Laforenza. 2006.
\newblock \href {https://doi.org/10.1145/1146847.1146881} {Query-driven
  document partitioning and collection selection}.
\newblock In \emph{Proceedings of the 1st International Conference on Scalable
  Information Systems}, InfoScale '06, page 34–es, New York, NY, USA.
  Association for Computing Machinery.

\bibitem[{Risvik et~al.(2013)Risvik, Chilimbi, Tan, Kalyanaraman, and
  Anderson}]{risvik2013maguro}
Knut~Magne Risvik, Trishul Chilimbi, Henry Tan, Karthik Kalyanaraman, and Chris
  Anderson. 2013.
\newblock Maguro, a system for indexing and searching over very large text
  collections.
\newblock In \emph{Proceedings of the sixth ACM international conference on Web
  search and data mining}, pages 727--736. ACM.

\bibitem[{Sasaki et~al.(2018)Sasaki, Sun, Schamoni, Duh, and
  Inui}]{sasaki-etal-2018-cross}
Shota Sasaki, Shuo Sun, Shigehiko Schamoni, Kevin Duh, and Kentaro Inui. 2018.
\newblock \href {https://doi.org/10.18653/v1/N18-2073} {Cross-lingual
  learning-to-rank with shared representations}.
\newblock In \emph{Proceedings of the 2018 Conference of the North {A}merican
  Chapter of the Association for Computational Linguistics: Human Language
  Technologies, Volume 2 (Short Papers)}, pages 458--463, New Orleans,
  Louisiana. Association for Computational Linguistics.

\bibitem[{Si and Callan(2003)}]{si_relevant_2003}
Luo Si and Jamie Callan. 2003.
\newblock \href {https://doi.org/10.1145/860435.860490} {Relevant document
  distribution estimation method for resource selection}.
\newblock In \emph{Proceedings of the 26th Annual International ACM SIGIR
  Conference on Research and Development in Informaion Retrieval}, SIGIR '03,
  page 298–305, New York, NY, USA. Association for Computing Machinery.

\bibitem[{Stratos(2019)}]{stratos_mutual_2019}
Karl Stratos. 2019.
\newblock \href {https://doi.org/10.18653/v1/N19-1113} {Mutual information
  maximization for simple and accurate part-of-speech induction}.
\newblock In \emph{Proceedings of the 2019 Conference of the North {A}merican
  Chapter of the Association for Computational Linguistics: Human Language
  Technologies, Volume 1 (Long and Short Papers)}, pages 1095--1104,
  Minneapolis, Minnesota. Association for Computational Linguistics.

\bibitem[{Stratos and Wiseman(2020)}]{stratos_learning_2020}
Karl Stratos and Sam Wiseman. 2020.
\newblock \href {http://proceedings.mlr.press/v119/stratos20a.html} {Learning
  discrete structured representations by adversarially maximizing mutual
  information}.
\newblock In \emph{Proceedings of the 37th International Conference on Machine
  Learning, {ICML} 2020, 13-18 July 2020, Virtual Event}, volume 119 of
  \emph{Proceedings of Machine Learning Research}, pages 9144--9154. {PMLR}.

\bibitem[{Thomas and Shokouhi(2009)}]{thomas_sushi_2009}
Paul Thomas and Milad Shokouhi. 2009.
\newblock \href {https://doi.org/10.1145/1571941.1572014} {Sushi: Scoring
  scaled samples for server selection}.
\newblock In \emph{Proceedings of the 32nd International ACM SIGIR Conference
  on Research and Development in Information Retrieval}, SIGIR '09, page
  419–426, New York, NY, USA. Association for Computing Machinery.

\bibitem[{{Tishby} et~al.(2000){Tishby}, {Pereira}, and
  {Bialek}}]{tishby_information_1999}
Naftali {Tishby}, Fernando C.~N. {Pereira}, and William {Bialek}. 2000.
\newblock The information bottleneck method.
\newblock \emph{Proc. 37th Annual Allerton Conference on Communications,
  Control and Computing, 1999}, pages 368--377.

\bibitem[{van Rijsbergen(1979)}]{van1979information}
C.~J. van Rijsbergen. 1979.
\newblock \emph{Information Retrieval}, 2ª edition.
\newblock Butterworth-Heinemann, London.

\end{thebibliography}

\newpage
\appendix
\vskip 2em \centerline{\Large \bf APPENDIX} \vskip -1em
\section{Experimental Setup}
\label{app:experimental_setup}
\subsection{Data Processing}

We use the \texttt{gensim} package to preprocess all the queries and documents into TF-IDF features in three steps: 1. Tokenize the sentences into words and remove stop words; 2. Create a dictionary with most frequent words (20,000 for ECSL, and 3,000 for CLIR); 3. Transform the queries and documents into TF-IDF vectors.

\begin{table}[!ht]
\centering
{\scriptsize
\begin{tabular}{|l|c|c|c|p{3.5em}|c|p{3.5em}|}
\hline
\multirow{2}{*}{Data sets} & \multicolumn{4}{c|}{Queries} & \multicolumn{2}{c|}{Documents} \\
  \cline{2-7}
 & \#Train & \#Dev & \#Test & Avg($|L|$) & \#Doc & Avg($|L|$) \\
  \hline
 ECSL & 339k & 38k & 55K & 4 & 4.7M & 16  \\ 
  \hline
 CLIR-FR & 14.8K & 4.9K & 4.9K & 21 & 240K & 177 \\ 
  \hline
 CLIR-IT & 14.8K & 4.9K & 4.9K & 21 & 198K & 169 \\ 
  \hline
 CLIR-SW & 12.8K & 4.1K & 4.1K & 18 & 24k & 82 \\
  \hline
 CLIR-TA & 14.3K & 4.7K & 4.7K & 19 & 27k & 79 \\
  \hline
\end{tabular}
}
\caption{This table shows the statistics of the ECSL data set and the CLIR data set, including the numbers of queries, documents and their average length.}
\label{table:datastat}
\end{table}

\subsection{Baselines}
\label{app:baselines}
Balanced K-means is built on top of K-means with approximated Hungarian algorithm to ensure the formed cluster sizes are well balanced. In K-means and Balanced K-means, documents and training queries are used together to form clusters, and we select the nearest clusters to the query vector as the top relevant clusters to route the query to. In IMSAT, the same model is used for both document assignment and query routing, after being trained on both queries and documents. KLD (Kullback-Liebler Divergence) only uses document information to form document clusters while QKLD (Query-based Kullback-Liebler Divergence) leverages the query information to form clusters. In KLD and QKLD, an additional shard selection algorithm is used to choose the relevant shards for new incoming queries.

\subsection{MICO}
During training, we set the batch size as $256$. We optimize the objective using \texttt{Adam} with learning rate $0.03$ for $\phi$ and $\psi$, and with learning rate $0.1$ for $\theta$. We model both $p_q^\psi(z|\bm{q})$ and $p_d^\phi(z|\bm{d})$ with neural networks with one hidden layer (dimension 20) and one softmax layer. We set the hyper-parameter for entropy regularization as $\beta=10$. To stablize the training process, we set gradient clipping as $10.0$. 

\subsection{MICO-q}
We set the hyper-parameters the same as in MICO except: 1. We set the gradient clipping as $1.0$; 2. We update $\theta$ for 4 steps instead of 1; 3. We set the entropy regularization strength as $\beta=3$; 4. We set the query consistency parameter as $\gamma=3$.

\subsection{Computing Environment}
We use an AWS EC2 instance (p3.2xlarge) consisting of 8 vCPUs (Intel Xeon E5-2686 v4) with 64GB memory, with one NVIDIA GPU card (Tesla V100-SXM2-16GB) for running our experiments. The computing environment is with Python 3.8, PyTorch 1.5, CUDA 9.2.

\section{Details of MICO}
\subsection{Preliminary}

We introduce the notations and the basic concept of mutual information briefly here.
We denote a random variable $X$'s probability function by $p_X(x)$. We shorten it to $p(x)$ or $p_X$ when its meaning is clear under its context. We use $p^\theta(x)$ when $p(x)$ is in a parametric distribution family and $\theta$ is the parameter.
$p_{XY}(x,y)$ is the joint distribution of two random variables $X$ and $Y$, and $p_{X|Y}(x|y)$ is the conditional distribution of $X$ given $Y$, while 
$p_Y(y)=\sum_{x} p_{XY}(x,y)$ is the marginal distribution of $Y$ under $p_{XY}(x,y)$. 
The expectation of a function $f(x)$ w.r.t. a random variable $x$ is defined by $\mathbb{E}_{x\sim p_X} [f(x)] = \sum_{x\in\mathcal{X}} f(x) p_X(x)$. 
We denote the entropy of $X$ by $H(X) = H(p_X) = \mathbb{E}_{x\sim p_X} [-\log (p_X(x))]$, 
while the conditional entropy is $H(Y|X) = \mathbb{E}_{x\sim p_X} H(p_{Y|X}) =\mathbb{E}_{(x,y)\sim p_{XY}}[-\log (p_{Y|X})]$.
The cross-entropy between $X$ and $X'$ is defined as $H(X;X') = \mathbb{E}_{x\sim p_X} [-\log (p_{X'}(x))]$. 
We denote the mutual information between $X$ and $Y$ by $I(X;Y)=\sum_x \sum_y p_{XY} \log\left( \frac{p_{XY}}{p_X p_Y}\right)=H(X)-H(X|Y)$. It is minimized at 0 when $X$ and $Y$ are independent, i.e., $p_{XY}=p_X p_Y$. The chain rule for mutual information is $I(X;(Y,Z))=I(X;Z)+I(X;Y|Z)$ where $I(X;Y|Z)=H(X|Z)-H(X|Y,Z)\geq 0$.

\subsection{Information Maximization Approximation}\label{sec:info_max_approx}
{We explain the reason of having the three components for calculating the mutual information.}
We seek to train the probabilistic modules $p^\psi(z|\bm{q})$ and $p^\phi(z'|\bm{d})$ through maximizing the mutual information between $Z$ and $Z'$ over our training data set, as shown in Figure \ref{fig:mico}. 
From this probabilistic graph, we know $Z$ and $Z'$ are independent when given $\bm{Q}$, i.e., $I(Z';Z|\bm{Q})=0$. Based on the chain rule of mutual information, we also have $I(Z';(\bm{Q},Z))=I(Z';\bm{Q})+I(Z';Z|\bm{Q})=I(Z';Z)+I(Z';\bm{Q}|Z)$. Therefore, $I(Z';Z) = I(Z';\bm{Q})-I(Z';\bm{Q}|Z)$. Since $I(Z';Z)$ is intractable under large amount of data and $I(Z';\bm{Q}|Z)\geq0$, we maximize $I(Z';\bm{Q})$ instead, which is an upper bound of $I(Z';Z)$. 
With the fact $I(Z';\bm{Q}) = H(Z') - H(Z'|\bm{Q})$, we need the marginal distribution $p^\phi(z')=\mathbb{E}_{\bm{d}\sim p_D}[p^\phi(z'|\bm{d})]$ and the conditional distribution $p^\phi(z'|\bm{q})=\sum_{\bm{d}}p^\phi(z'|\bm{d})p(\bm{d}|\bm{q})$, both of which are also part of the logarithm part of the two entropy terms. Therefore, $I(Z';\bm{Q})$ is again intractable if we only use $p^\phi(z'|\bm{d})$. We approximate $p^\phi(z')$ and $p^\phi(z'|\bm{q})$ by neural networks $g^\theta(z')$ and $p^\psi(z|\bm{q})$ respectively, which falls into the category of variational methods. As $H(Z') = H(p^\phi(z')) \leq H(p^\phi(z');g^\theta(z'))$ and $H(Z'|\bm{Q}) = \mathbb{E}_{\bm{q}\sim p_Q}H(p^\phi(z'|\bm{q})) \leq \mathbb{E}_{\bm{q}\sim p_Q}H(p^\phi(z'|\bm{q});p^\psi(z|\bm{q}))$, we upper-bound
$H(Z'|\bm{Q})$ with $H(Z'; Z) := \mathbb{E}_{(\bm{q},\bm{d})\sim p_{QD}}[H(p^\phi(z'|\bm{d});p^\psi(z|\bm{q}))]$,
and introduce 
$H^+(Z') := \mathbb{E}_{\bm{d}\sim p_D}H(p^\phi(z'|\bm{d});g^\theta(z'))$ as an upper bound of $H(Z')$. Note that with $H(X) = \min_{Y}H(X;Y)$, our approximation $\hat{I}(Z';\bm{Q})=H^+(Z') - H(Z';Z)$ gives exact evaluation of $I(Z';\bm{Q})$ when $H^+(Z')$ is minimized w.r.t. $\theta$ when $g^\theta(z')=p^\phi(z')$, and $H(Z';Z)$ is minimized w.r.t. $\psi$ when $p^\psi(z'|\bm{q})=p^\phi(z'|\bm{q})$.

\section{Theoretical Analysis of Stochastic Optimization of MICO}
\label{app:sgd_mico}

First we show why computing $I(Z'; Z)$ is intractable even under the stochastic optimization framework. With the MICO framework, the expectation over $p_{QD}$ is approximated by the average over one batch of data $\bm{B}=\{(\bm{q}_1, \bm{d}_1), (\bm{q}_2, \bm{d}_2), \dots, (\bm{q}_b, \bm{d}_b)\}$. If one objective is $f$ consisting of the expectation over $p_{QD}$, we evaluate $f$ on $\bm{B}$ and denote the corresponding result as $f_{\bm{B}}$. The core assumption in stochastic optimization is $f = \mathbb{E}_{\bm{B}}f_{\bm{B}}$. 

For 
\begin{align*}
&I(Z'; Z) = \underset{\substack{(\bm{q},\bm{d})\sim p_{QD}}}{\bbE} \left[\sum_{z,z'} p^\phi(z'|\bm{d}) p^\psi(z|\bm{q}) \log \right.\\
&\left.\left(\frac{\mathbb{E}_{(\bm{q},\bm{d})\sim p_{QD}} [p^\phi(z'|\bm{d}) p^\psi(z|\bm{q})]}{\mathbb{E}_{(\bm{q},\bm{d})\sim p_{QD}}[p^\phi(z'|\bm{d})] \mathbb{E}_{(\bm{q},\bm{d})\sim p_{Q,D}} [p^\psi(z|\bm{q})]}\right)\right],
\end{align*}
if we use
\begin{align*}
&I_{\bm{B}}(Z'; Z) = \frac{1}{b}\sum_{i=1}^b \left[\sum_{z,z'} p^\phi(z'|\bm{d}_i) p^\psi(z|\bm{q}_i) \log\right.\\
&\left.\left( \frac{\frac{1}{b}\sum_{i=1}^b [p^\phi(z'|\bm{d}_i) p^\psi(z|\bm{q}_i)]}{\frac{1}{b}\sum_{i=1}^b[p^\phi(z'|\bm{d}_i)] \frac{1}{b}\sum_{i=1}^b [p^\psi(z|\bm{q}_i)]}\right)\right],
\end{align*}
it violates the assumption since $I(Z'; Z) \neq \mathbb{E}_{\bm{B}}[I_{\bm{B}}(Z'; Z)]$.

In MICO, we replace $I(Z'; Z)$ with $I(Z';\bm{Q})$.

Moreover, we use
\begin{align*}
&{H}_{\bm{B}}(Z';Z) = \frac{1}{b}\sum_{i=1}^b \left[\sum_{z=z'}- p^\phi(z'|\bm{d}_i) \log\right.\\
&\left.\left( p^\psi(z|\bm{q}_i)\right)\right]
\end{align*}
and 
\begin{align*}
&{H}_{\bm{B}}(Z') = \frac{1}{b}\sum_{i=1}^b \sum_{z'}- p^\phi(z'|\bm{d}) \log\\
&\left( \frac{1}{b}\sum_{i=1}^b p^\phi(z'|\bm{d}_i)\right)
\end{align*}
to approximate
\begin{align*}
&{H}(Z';Z) = \mathbb{E}_{(\bm{q},\bm{d})\sim p_{Q,D}} \left[\sum_{z=z'}- p^\phi(z'|\bm{d}) \log \right.\\
&\left.\left( p^\psi(z|\bm{q})\right)\right]
\end{align*}
and
\begin{align*}
&{H}(Z') = \sum_{z'}-\mathbb{E}_{(\bm{q},\bm{d})\sim p_{Q,D}}[p^\phi(z'|\bm{d})] \log\\
&\left(\mathbb{E}_{(\bm{q},\bm{d})\sim p_{Q,D}}[p^\phi(z'|\bm{d})]\right),
\end{align*}
respectively.

Note that ${H}(Z',Z) = \mathbb{E}_{\bm{B}}[{H}_{\bm{B}}(Z',Z)]$ while ${H}(Z') \neq \mathbb{E}_{\bm{B}}[{H}_{\bm{B}}(Z')]$. This is due to the logarithm part in $H_{\bm{B}}(Z')$. Without using the stochastic approximation on the logarithm, since it is expensive to evaluate the expectation of $p^\phi(z'|\bm{d})$ on the full dataset, we have to replace it with $g^\theta(z')$ in the logarithm term which creates an upper bound ${H}^+(Z')$ on ${H}(Z')$ as an approximation (used in the tractable loss function of MICO):
\begin{align*}
&{H}^+(Z') = \sum_{z'}-\mathbb{E}_{(\bm{q},\bm{d})\sim p_{Q, D}}[p^\phi(z'|\bm{d})] \log\left(g^\theta(z')\right). \\
&{H}_{\bm{B}}^+(Z') = \frac{1}{b}\sum_{i=1}^b \sum_{z'}- p^\phi(z'|\bm{d}) \log\left( g^\theta(z')\right), \\
&{H}^+(Z') = \mathbb{E}_{\bm{B}}[{H}_{\bm{B}}^+(Z')]
\end{align*}
The stochastic evaluation of $\mathcal{L}$ is then $\mathcal{L}_{\bm{B}} = \beta{H}_{\bm{B}}^+(Z')-{H}_{\bm{B}}(Z';Z)$ and we have $\mathcal{L} = \mathbb{E}_{\bm{B}}[\mathcal{L}_{\bm{B}}]$, which is also the reason of why we use $H(Z';Z)$ instead of $H(Z'|\bm{Q})$ in MICO.

For each batch of data, since ${H}(Z') = \min_{g^\theta(z')} {H}^+(Z')$, we first update $g^\theta(z')$ for one step to descend $H^+_{\bm{B}}(Z')$ (corresponding to SGD), and then update $p^\psi(z'|\bm{q})$ and $p^\phi(z'|\bm{d})$ for one step to descend $\mathcal{L}_{\bm{B}}$ (corresponding to another SGD).

\end{document}